\documentclass[a4paper,fleqn,usenatbib]{mnras}

\usepackage{newtxtext,newtxmath}
\usepackage[T1]{fontenc}
\usepackage{ae,aecompl}

\usepackage{graphicx}	% Including figure files
\usepackage{amsmath}	% Advanced maths commands
\usepackage{amssymb}	% Extra maths symbols

\newcommand{\gaia}{\emph{Gaia}}
\newcommand{\MG}{$M_G$}
\newcommand{\DMG}{$\Delta M_G$}

\newcommand{\Ks}{$K_s$}
\newcommand{\Kserr}{$\sigma_{K_s}$}

\newcommand{\GKd}{$(G-K_s)_0$}
\newcommand{\teff}{$T_{\rm eff}$}
\newcommand{\logg}{$\log g$}
\newcommand{\feh}{[Fe/H]}

\newcommand{\fbin}{$f_{b}$}

\newcommand{\mpri}{$m_1$}

\newcommand{\Mini}{$M_{ini}$}
\newcommand{\Msun}{$M_\odot$}
\newcommand{\parallax}{$\varpi$}
\newcommand{\parerr}{$\sigma_\varpi$}
\newcommand{\bpflux}{\texttt{phot\_bp\_mean\_flux}}
\newcommand{\rpflux}{\texttt{phot\_rp\_mean\_flux}}

\newcommand{\bpfluxerr}{\texttt{phot\_bp\_mean\_flux\_error}}
\newcommand{\rpfluxerr}{\texttt{phot\_rp\_mean\_flux\_error}}
\newcommand{\gflux}{\texttt{phot\_g\_mean\_flux}}
\newcommand{\gfluxerr}{\texttt{phot\_g\_mean\_flux\_error}}
\newcommand{\glow}{low-$\gamma$}
\newcommand{\ghigh}{high-$\gamma$}

\title[Solar-type Field Binary Stars]{Smoking Gun of the Dynamical Processing of the Solar-type Field Binary Stars}

\author[C. Liu]{
Chao Liu$^{1,2}$\thanks{E-mail: liuchao@nao.cas.cn}
\\
$^{1}$Key Lab of Optical Astronomy, National Astronomical Observatories, CAS, Beijing 100101, China\\
$^{2}$University of Chinese Academy of Sciences, Beijing 100049, China
}

\date{Accepted XXX. Received YYY; in original form ZZZ}

\pubyear{2019}

\begin{document}
\label{firstpage}
\pagerange{\pageref{firstpage}--\pageref{lastpage}}
\maketitle

\begin{abstract}
We investigate the binarity properties in field stars using more than 50\,000 main-sequence stars with stellar mass from 0.4 to 0.85\,$M_\odot$ observed by LAMOST and {\emph Gaia} in the solar neighborhood. By adopting a power-law shape of the mass-ratio distribution with power index of $\gamma$, we conduct a hierarchical Bayesian model to derive the binary fraction ($f_{b}$) and $\gamma$ for stellar populations with different metallicities and primary masses ($m_1$).
We find that $f_b$ is tightly anti-correlated with $\gamma$, i.e. the populations with smaller binary fraction contains more binaries with larger mass-ratio and vice versa.
The high-$\gamma$ populations with $\gamma>1.2$ have lower stellar mass and higher metallicity, while the low-$\gamma$ populations with $\gamma<1.2$ have larger mass or lower metallicity.  The $f_b$ of the high-$\gamma$ group is anti-correlated with [Fe/H] but flat with $m_1$. 
Meanwhile, the $f_b$ of the low-$\gamma$ group displays clear correlation with $m_1$ but quite flat with [Fe/H]. 
The substantial differences are likely due to the dynamical processing when the binaries were in the embedded star clusters in their early days.
The dynamical processing tends to destroy binaries with smaller primary mass, smaller mass-ratio, and wider separation. Consequently, the
 high-$\gamma$ group containing smaller $m_1$ is more effectively influenced and hence contains less binaries, many of which have larger mass-ratio and shorter period. However,  the low-$\gamma$ group is less affected by the dynamical processing due to their larger $m_1$. These are evident that the dynamical processing does effectively work and significantly reshape the present-day binary properties of field stars.
\end{abstract}

\begin{keywords}
binaries: general -- Hertzsprung-Russell and colour-magnitude diagrams -- Stars: solar-type -- Stars: abundances -- Stats:formation -- methods:statistical
\end{keywords}

%%%%%%%%%%%%%%%%%%%%%%%%%%%%%%%%%%%%%%%%%%%%%%%%%%

%%%%%%%%%%%%%%%%% BODY OF PAPER %%%%%%%%%%%%%%%%%%

\section{Introduction}\label{sec:intro}
%%%%%%%%%%%%%%%%%%%%%%%%%%%%%%%%%%%%%%%%%%%%%%%%%%%%%%

Binary stars are very common in stellar systems with different scales, from star cluster to the whole galaxy. Some literatures even thought that the most of stars live in binary systems~\citep{Heintz1969,Abt1976,Duquennoy1991}. The statistical analysis of binary systems play important role to constrain star formation in different environments. The statistical properties that can well describe binary systems include the fraction of binary, the distribution of mass-ratio, the distribution of orbital period, and the distribution of the orbital eccentricity. These properties are usually compared between populations with different age, metallicity, and primary mass~\citep{Duquennoy1991,Henry1990,Fischer1992,Mason1998a,Mason1998b}. 

The association of binary fraction with metallicity has been broadly studied by lots of observational~\citep*[][etc.]{Carney1983,Latham2002,Raghaven2010,Rastegaev2010,Moe2013,Gao2014,Hettinger2015,Gao2017,Tian2018,Badenes2018,Moe2018} and theoretical works~\citep[][etc.]{Machida2009,Kratter2010,Myers2011,Bate2014,Tanaka2014}. Recently, lots of works, especially depending on large volume of datasets, found that binary fraction, at least for close binaries, is anti-correlated with metallicity~\citep[e.g.][]{Raghaven2010,Gao2014, Gao2017, Tian2018, Badenes2018, Moe2018}, while a few others claimed that \fbin\ is positively correlated~\citep[e.g.][]{Carney1983,Abt1987,Hettinger2015} or not correlated with metallicity~\citep{Latham2002,Moe2013}. 

Among these works, \citet{Moe2018} reanalyzed five different datasets of close binaries with sample size from several hundreds to several ten thousands from literatures and concluded that the binary fraction is strongly anti-correlated with metallicity. The datasets used in their studies contain close binaries with separation smaller than $10$\,AU. As complementary, \citet{ElBadry2019} discussed about the variation of the binary fraction as a function of metallicity for wide binaries with separation of $50$--$50\,000$\,AU. They found the anti-correlation between the binary fraction and metallicity exists only when the separation of the two companions is between $50$ and $100$\,AU. It becomes flat for binaries with larger separations as $100<a<200$\,AU.

The binary fraction of main sequence (MS) stars is also found positively correlated with primary mass in many works~\citep[see the review by][and references therein]{Duchene2013}. For very low-mass field binaries (\mpri$<0.1$\,\Msun), the binary fraction is only about $22$\%~\citep{Allen2007}. For low-mass binaries ($0.1<m_1<0.5$\,\Msun), the fraction increases to around $26$\%~\citep{Delfosse2004,Dieterich2012}. For solar-type stars ($0.5<m_1<1.4$\,\Msun, which is also defined as the MS stars with luminosity from $0.1$ to $10$\,$L_\odot$ by~\citet{Raghaven2010}), the fraction further increases to around $41$--$50$\%~\citep{Raghaven2010}. And for the intermediate-mass ($1.5<m_1<5$\,\Msun) and massive stars ($m_1>16$\,\Msun), the fraction is larger than $50$\%~\citep{Chini2012,Mason2009,Sana2013}. 

%Close binary fraction increases with primary mass (Abt et al. 1990; Raghavan et al. 2010; Sana et al. 2012; Duchene \& Kraus 2013; Moe \& Di Stefano 2017; Murphy et al. 2018).

For the well studied solar-type binaries, the arguments about mass-ratio distribution are contradicted. ~\citet{Duquennoy1991} derived a mass-ratio distribution biased to lower mass-ratio for solar-type stars, which makes it very similar to the canonical initial mass function (IMF)~\citep{Miller1979,Kroupa1990}. However, later studies claimed different shape. \citet{Raghaven2010} found that the mass-ratio distribution is rather flat with a peak at $q>0.9$, which is not like any of the canonical IMF~\citep*{Bastian2010}. \citet*{Goldberg2003}, on the other hand, showed a double peaked mass-ratio distribution. \citet{Duchene2013} compiled many references and found that if mass-ratio distribution is described with a power law, then the power index decreases with primary mass. Again, they demonstrated that the mass-ratio distribution, which is equivalent with the present-day mass function of the secondary, is significantly biased from any of the canonical IMF.  

A large fraction of previous studies on the binary fraction and mass-ratio distribution were based on small dataset with number of stars from a dozen to a few hundreds, which suffers from substantially large Poisson noise. In recent years, some of the works~\citep{Gao2014, Gao2017,Hettinger2015, Badenes2018, Moe2018, ElBadry2018b,ElBadry2019} have made use of large survey data. 

In light of these studies, we make the measurement of the binary fraction and mass-ratio distribution for stars with different metallicity and primary mass using the LAMOST data~\citep{Zhao2012} combined with~\gaia\ DR2~\citep{gaia2016,Gaia2018}. LAMOST DR5 have collected more than 8 million stellar spectra with spectral resolution of $R=1800$ and limiting magnitude of $r=18$\,mag~\citep{Deng2012}. It provides stellar effective temperature, surface gravity, and \feh, with precision of about 110\,K, 0.2\,dex, and 0.15\,dex~\citep{Luo2015,Gao2015}.  These parameters enable us to derive the stellar mass for solar-type stars. On the other hand, \gaia\ DR2 provides very accurate photometries in $G$, $BP$, and $RP$ bands as well as astrometry \citep{Gaia2018,Luri2018,Lindegren2018}. With accurate luminosity derived from parallax, field stars can be mapped in the Hertzprung-Russell (HR) diagram with clear binary sequence located around 0.75\,mag above the single main-sequence \citep[see][]{HRD2018}. Hence, one can study the statistical properties of binaries directly in HR diagram, similar to the binary studies based on the photometries of star clusters or OB associations~\citep*[e.g.][]{Kouwenhoven2007,Milone2012,Li2013,Yang2018}.

The paper is organized as follows. Section~\ref{sec:data} describe how we select the solar-type stars and how to estimate their parameters. Section~\ref{sec:method} develops the hierarchical Bayesian model which can simultaneously derive binary fraction and mass-ratio distribution. Section~\ref{sec:result} shows the results for stellar populations with various metallicity and primary mass. From the Bayesian model, we also provide probabilities to be binary of the sample stars with the mass-ratio estimates. Section~\ref{sec:disc} raises discussions about the caveats of the method and comparisons with other works. Finally, brief conclusions are drawn in section~\ref{sec:conclusion}.

%%%%%%%%%%%%%%%%%%%%%%%%%%%%%%%%%%%%%%%%%%%%%%%%%%%%%%     
\section{Data}\label{sec:data}
%%%%%%%%%%%%%%%%%%%%%%%%%%%%%%%%%%%%%%%%%%%%%%%%%%%%%%

%=====================================================
\subsection{Data selection}\label{sec:dataselect}
%=====================================================

We select common stars cross-identified from LAMOST DR5 (\url{http://dr5.lamost.org}), \gaia\ DR2~\citep{Gaia2018}, and 2MASS~\citep{2mass}. The cross-identification is conducted by comparing the right ascensions and declinations in equinox of J2000.0 among the three catalogs. Because the diameter of the LAMOST fiber is $3.3$\,arcsec and the mean seeing during LAMOST observations is around $3$ \,arcsec (Shi, J-R. private communication), the spatial resolution of LAMOST should be around $5$\,arcsec, which is used as the radius of circle in the cross-identification. Only the star pairs with nearest positions within the circle are selected. 

First, we derive absolute magnitude in $G$-band of the stars from \gaia\ parallax based on the likelihood distribution. which can be written as
\begin{equation}\label{eq:MG}
p(M_G'|\varpi,G) \propto \exp\left(-\frac{(10^{-(G-M_G'+5)/5}-\varpi)^2}{2\sigma_\varpi^2}\right),
\end{equation}
where $M_G'$ is the \gaia\ $G$-band absolute magnitude without extinction correction, $\varpi$ and $\sigma_\varpi$ are the parallax and its uncertainty, respectively. $M_G'$ equals to $ M_G+A_G$, where $M_G$ and $A_G$ are the \gaia\ $G$-band absolute magnitude with extinction correction and the interstellar extinction in $G$-band, respectively. We estimate the maximum likely $M_G'$ and is uncertainty for each star by arbitrarily drawing 10\,000 points from the distribution defined by Eq.~(\ref{eq:MG}).

We then select samples using the following criteria:
\begin{enumerate}
	\item[1]$M_G'>4$\,mag;
	\item[2]\parallax$>3$ so that most of the stars are located within $\sim333$\,pc;
	\item[3]\parallax$/$\parerr$>5$;
	\item[4]\bpflux$/$\bpfluxerr$>20$, \rpflux$/$\rpfluxerr$>20$, and \\\gflux$/$\gfluxerr$>20$;
	\item[5]$\sqrt{\chi^2/(\nu-5)}<1.2\times\max(1,\exp(-0.2(G-19.5)))$, where $\chi^2$ and $\nu$ represent for \texttt{astrometric\_chi2\_al} and \texttt{astrometric\_n\_good\_obs\_al}, respectively;
	\item[6]\Ks$<14.3$\,mag and \Kserr$<0.2$\,mag;
	\item[7] \teff\ derived from LAMOST is in between $3800$ and $<6500$\,K;\\
	\item[8] signal-to-noise ratio at $g$ band of LAMOST spectra is larger than 20.
\end{enumerate}
%\item[5](\bpflux$+$\rpflux)$/$\\\gflux$\times(1.2+\\0.03($\bpmag$-$\rpmag$)^2)<1.2$;

The first criterion ensures that the hot MS stars are removed.% \teff$>3800$\,K is subject to the lower limit of effective temperature estimates in LAMOST pipeline. We set the upper limit of \teff\ at $6500$\,K so that the turn-off stars are excluded from this sample. 
The second criterion selects the stars located within $\sim333$\,pc so that the data comply with the requirement of the completeness. %That is, within about $333$\,pc, the range of the %absolute magnitude and the initial mass of primary stars should be about complete, as seen in Figure~\ref{fig:complete}, in which the distributions of stars in \MG\ and stellar mass along distance is shown.
The faintest apparent magnitude for a star with mass of $0.08$\,\Msun\ at distance of $333$\,pc is between $14.5$ and $16.1$\,mag depending on different metallicity. This is bright enough for the companion in a binary to contribute luminosity that is able to be detected by LAMOST. Therefore, the volume cut enables the detection of the faintest secondaries.
The criteria 3--6 is set to select stars with accurate photometry and astrometry. Among them, criterion 5 is adopted from \citet{ElBadry2018c}. Criteria 7 and 8 help to select high quality spectra from LAMOST so that the stellar parameters derived from the spectra are reliable. Meanwhile, the two criteria also comply with the criterion 2.
After applying these selection criteria, we obtain 101\,061 MS stars in total. 

%=====================================================
\subsection{Stellar mass}\label{sec:distance}
%=====================================================

We adopt the present-day mass as the initial mass for the primary stars. First, because in this work we only focus on the main-sequence stars with stellar mass of 0.4-1\,\Msun, the mass loss due to stellar wind is very small and ignorable. Second, the fraction of mass transferring binary system is less than one percent~\citep{Rucinski1994PASP..106..462R,Chen2016} for the solar-mass main-sequence stars, hence the change of the stellar mass due to mass transfer is also ignored. 

The initial stellar mass of each star is determined by comparing their \teff, \logg, and \feh\ with PARSEC isochrones \citep{Bressan2012} by means of likelihood (see the details in Appendix~\ref{sec:mass}). 

Although the stellar parameters obtained from the LAMOST spectra are dominated by the primary stars in a binary system, the secondary stars may averagely reduce the effective temperature by $\sim50$\,K when $q>0.6$ and induce additional 0.1\,dex and 0.05\,dex dispersions in \logg\ and \feh, respectively, for the case of LAMOST spectra~\citep{ElBadry2018a}. Considering these systematics and uncertainties, the total uncertainty of the likelihood approach is about $0.05$\,\Msun.

%=====================================================
\subsection{Interstellar extinction correction}\label{sec:extinction}
%=====================================================

The determination of the interstellar extinction of the stars is described in Appendix~\ref{sect:extinction}. The uncertainty of the extinction in $G$ band is around $0.05$\,mag.

We further select the stars with $1.5<(G-K_s)_0<2.6$, where $(G-K_s)_0$ is the dereddened color index, so that the data can cover stellar mass from $0.4$ to $1.0$\,\Msun. A few stars without initial mass estimates due to their unusual photometries are excluded. After this cut, the samples are reduced to 76\,507 stars.  

\begin{figure}
	\centering
	\includegraphics[scale=0.43]{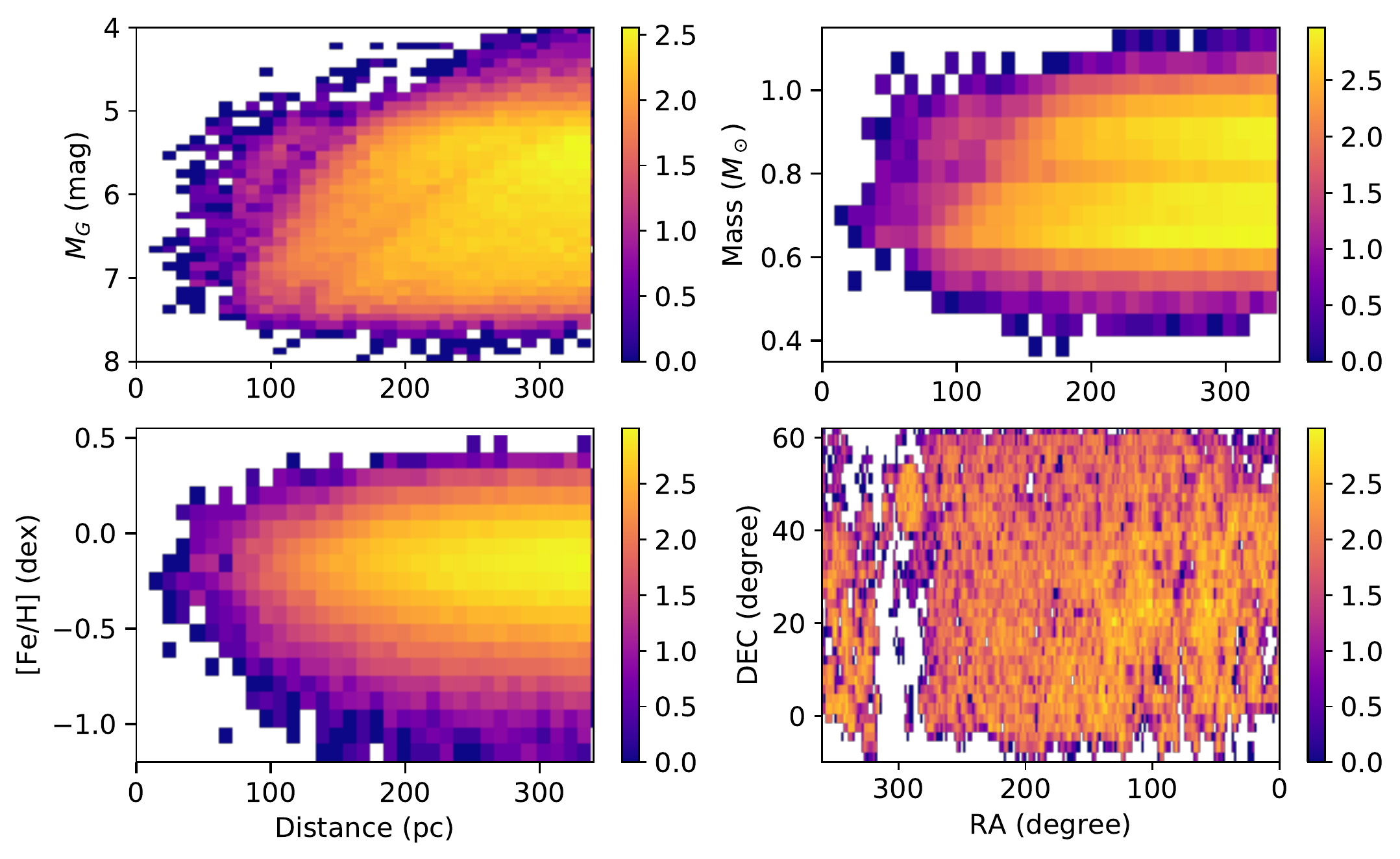}
	\caption{The top-left panel displays the distribution of the sample stars in \MG\ vs. distance plane. The top-right panel shows the distribution of the stars in stellar mass vs. distance plane. The bottom-left panel shows the distribution of \feh\ along distance. The bottom-right panel shows the distribution in right ascension and declination of the samples. The colors indicate the logarithmic density of stars in the bins.}\label{fig:complete}
\end{figure}

%=====================================================
\subsection{Spatial distribution}\label{sec:finalsample}
%=====================================================

We adopt the Bayesian distance derived from the parallax of \emph{Gaia} by \citet{BailerJones2018}. The top panels of Figure~\ref{fig:complete} demonstrate the distributions of the stars in \MG\ and mass vs. distance planes. It shows that \MG\ covering from 4 to 8\,mag does not show correlation with distance from 50 to 333\,pc. Meanwhile, stellar mass in the range from 0.4 to 1\,\Msun\ is also independent on distance. The bottom-left panel of the figure shows that \feh\ is uncorrelated with distance. These mean that \MG, stellar mass, and \feh\ of the samples are unbiased within the volume from 50 to 333\,pc. 

Although LAMOST survey only focuses on northern sky, it basically continuously cover the sky of about 20\,000 square degrees, from low to high Galactic latitude, as seen in the bottom-right panel of Figure~\ref{fig:complete}. Therefore, the samples selected in this work is representative to the stellar populations in the solar neighborhood.  

%=====================================================
\subsection{Color-absolute magnitude diagrams of the stars of different \feh}
%=====================================================

\begin{figure*}
	\centering
	\includegraphics[scale=0.5]{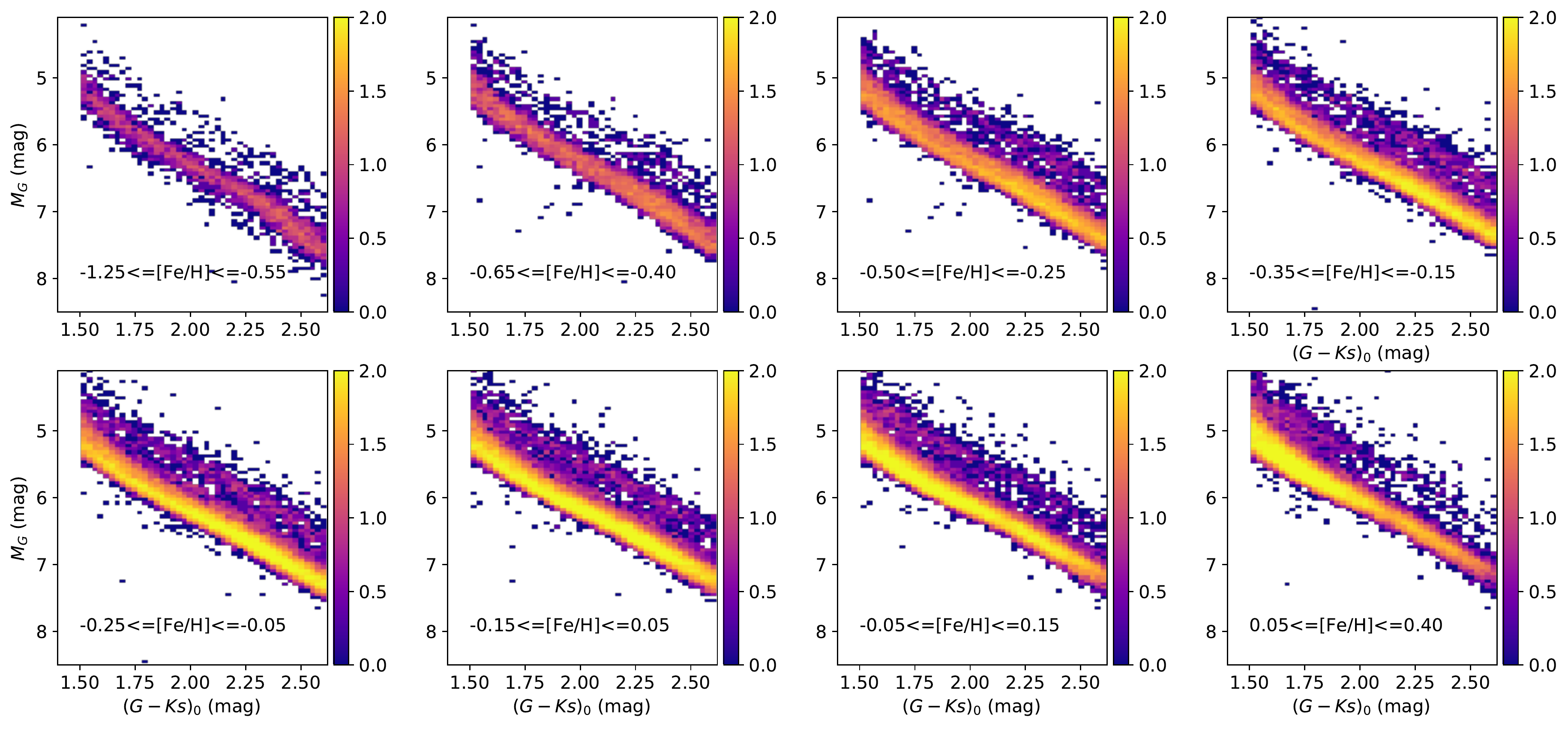}
	\caption{The color coded logarithmic density distributions of the samples in \MG\ vs. \GKd\ plane with different \feh\ bins, which ranges are indicated in the bottom of the panels. The reddening of the stars have been corrected in color indices.}\label{fig:cmd_feh}
\end{figure*}
Because different \feh\ may shift the main-sequence up and down in HR diagram, they can blur out the binary sequence. Therefore, we separate the stars in different \feh\ bins. The uncertainty of the LAMOST derived \feh\ is around 0.15\,dex~\citep{Gao2015}. Thus, we set the bin size of \feh\ lager than 0.2\,dex. On the other hand, we need to keep sufficient number of stars in each \feh\ bin for further separation in primary mass in section~\ref{sec:result}. Therefore, we adopt 8 \feh\ bins with ranges of $-1.25<$\feh$<-0.55$, $-0.65<$\feh$<-0.40$, $-0.50<$\feh$<-0.25$, $-0.35<$\feh$<-0.15$, $-0.25<$\feh$<-0.05$, $-0.15<$\feh$<+0.05$, $-0.05<$\feh$<+0.15$, and $+0.05<$\feh$<+0.40$. Note that the neighboring bins are partly overlapped with each other such that the number of stars in each bin is as large as possible. The overlapping would not harm the final results, but only smooth the results to some extent.
The density distributions of the sample stars in \MG\ vs. \GKd\ plane with different \feh\ bins is displayed in Figure~\ref{fig:cmd_feh}.
The density map is calculated in the grid of \GKd\ and \MG\ with size of $0.05\times0.025$\,mag. The colors in the figure codes the logarithmic density of stars. It is clearly seen two sequences in the color-absolute magnitude diagrams such that the unresolved binary sequence is located at upper side of the single star sequence due to their larger luminosity contributed by both companions. 

The shift of the absolute magnitude of an unresolved binary star from the single sequence reflects the mass-ratio of the binary system, which is denoted as $q$$=m_2/m_1$, where $m_1$ and $m_2$ are the masses of the primary and secondary stars, respectively. When $q$ is small ($\lesssim0.7$), the contribution to the luminosity from the secondary star is only around 0.1\,mag. Therefore, the composed luminosity for such binaries is similar to the single stars. When $q$ becomes larger ($\gtrsim0.7$), the contribution from the secondary becomes obvious, hence the luminosity of the whole system increases substantially. When $q$ is close to 1, the composed luminosity is doubled, which is equivalent to 0.75\,mag brighter.

\begin{figure}
	\includegraphics[scale=0.5]{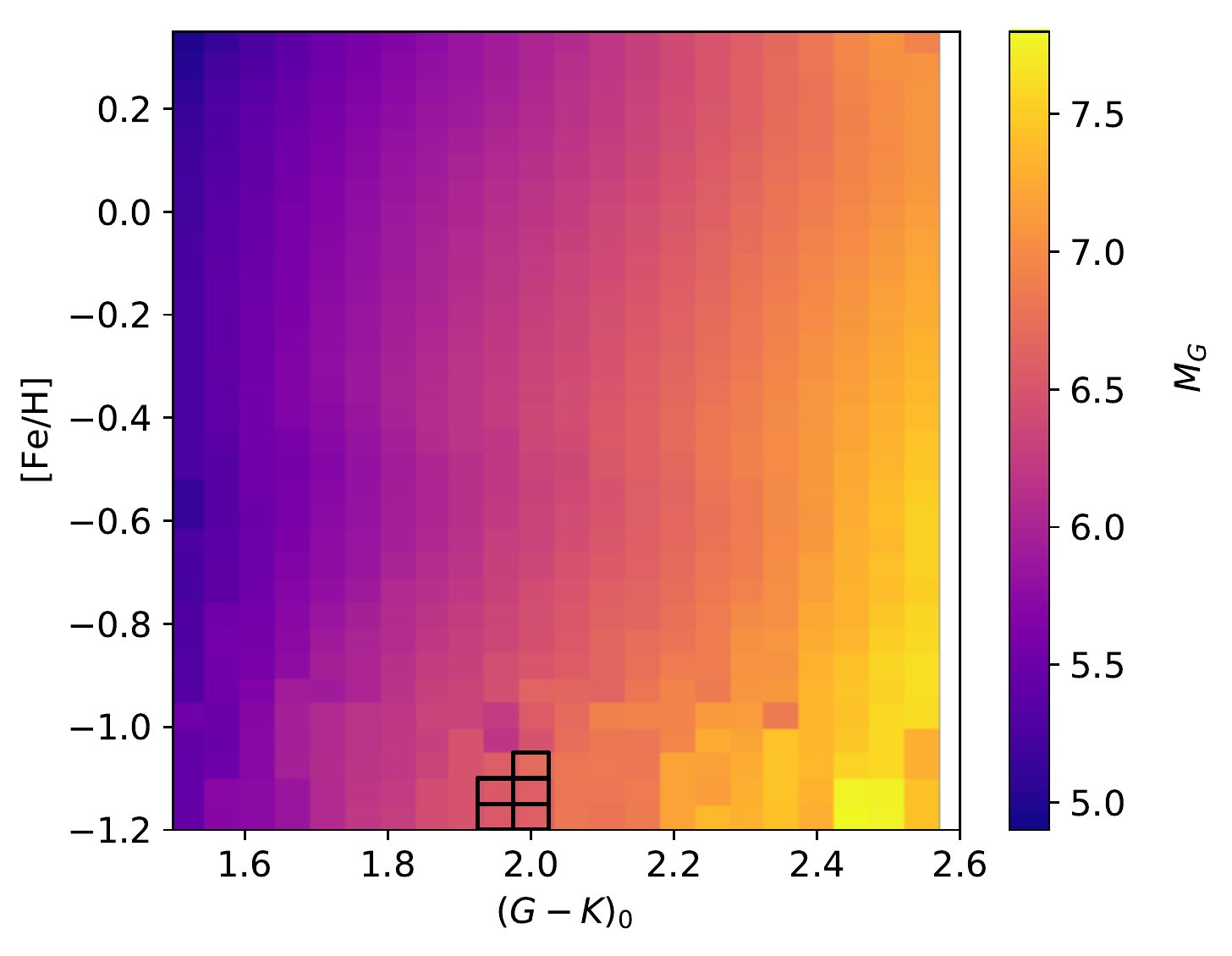}
	\caption{The colors show the intrinsic absolute magnitude of single stars in $G$ band as a function of \GKd\ and \feh. The five points marked with black frames are interpolated from the neighboring points.}\label{fig:MGoffset}
\end{figure}
To quantify the difference between the single and binary stars in \MG, we firstly address the ridge lines of the single MS stars in a finer \feh\ grid with bin size of $0.05$\,dex. For each \feh\ slice, we further separate the stars into \GKd\ bins with size of $0.05$\,mag. At each color index bin, we apply a kernel density estimation (KDE) to produce the density distribution of \MG. Specifically, the density of \MG\ is
\begin{equation}
p(M_G|[Fe/H], (G-K)_0)=\sum_{i=1}^{n}{K(\frac{M_G-M_{G,i}}{\sigma_{M_{G,i}}})},	
\end{equation}
where $K(t)=\exp(-t^2/2)$ is the kernel function; $M_{G,i}$ and $\sigma_{M_{G,i}}$ are the absolute magnitude and its uncertainty, respectively, of the $i$th star. We then locate the intrinsic \MG\ of single stars at the peak of $p(M_G|[Fe/H], (G-K)_0)$. Figure~\ref{fig:MGoffset} shows the color-coded intrinsic $G$-band absolute magnitude, denoted as $M_{G,single}([Fe/H],(G-K)_0)$, of the single star in \feh\ vs. \GKd\ plane. Note that there five points marked as black frames in the figure are failed to derive $M_{G,single}([Fe/H],(G-K)_0)$ due to the missing data. We then provide the linear interpolated values for the five points based on their neighboring points.

For a given star with fixed \feh\ and \GKd, we then calculate the differential absolute magnitude $\Delta M_G=M_G-M_{G,single}([Fe/H],(G-K)_0)$, in which $M_{G,single}$ is selected from Figure~\ref{fig:MGoffset} with nearest \feh\ and \GKd. The uncertainty of \DMG\ is contributed from several ways. First, it is mostly from the astrometric and photometric uncertainties. Then, the de-reddening process also contributes an additional uncertainty. Finally, $M_{G,single}$ also brings some uncertainty. Combining all these together, we obtain the typical uncertainty of \DMG\ as $0.08$\,mag.

\begin{figure*}
	\centering
	\includegraphics[scale=0.5]{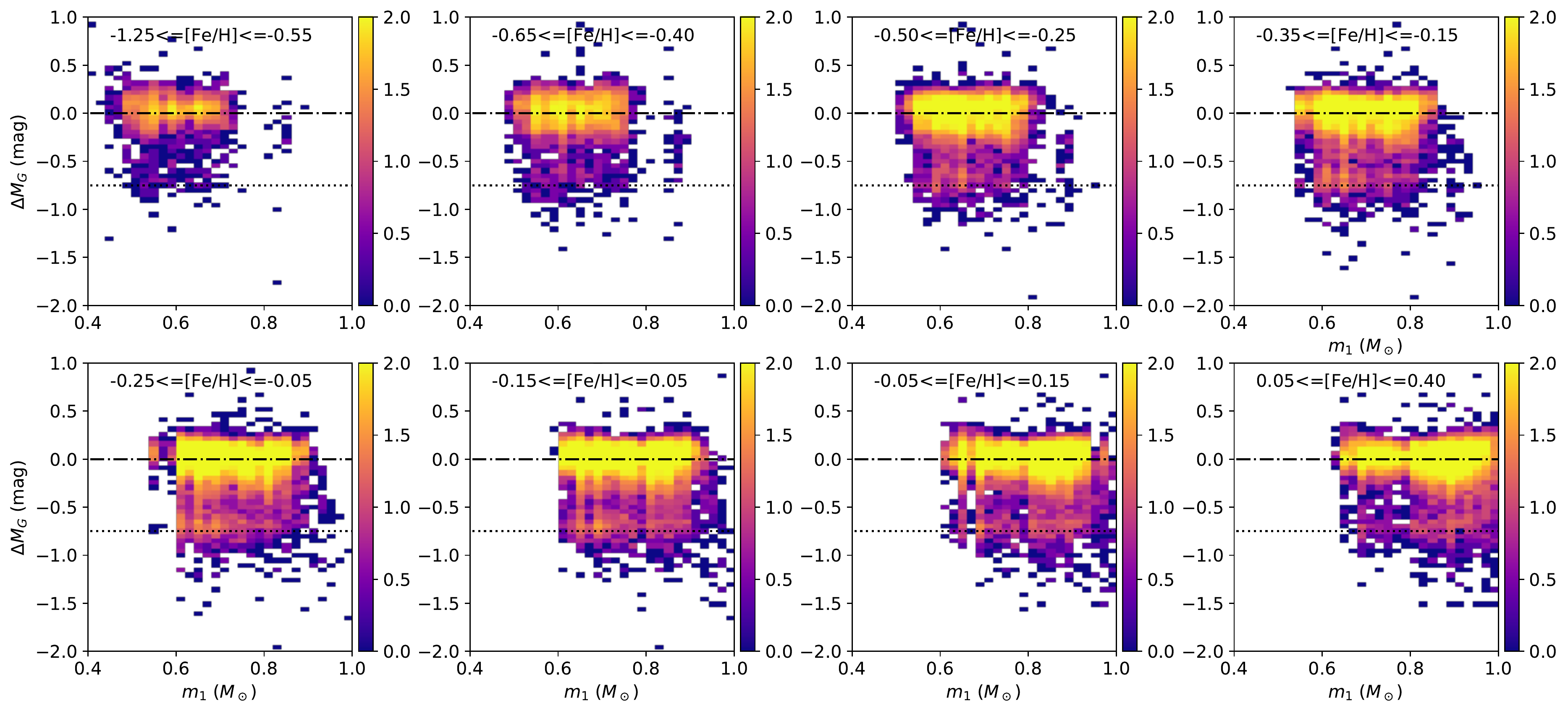}
	\caption{The color coded density distributions of the stars in \DMG\ vs. \mpri\ plane for various \feh\ bins. The horizontal dot-dashed lines in the panels indicate \DMG$=0$, which is the location of single stars. The horizontal dotted lines at \DMG$=-0.75$\,mag indicate the location of binary stars with mass-ratio of 1.}\label{fig:m1MG_feh}
\end{figure*}

Figure~\ref{fig:m1MG_feh} shows the distributions of the stars in \DMG\ vs.\ \mpri\ plane for different \feh\ populations. It is seen that a large fraction of the stars are aligned with \DMG$=0$, which is the location of the single stars. The vertical dispersion of these stars is roughly consistent with the uncertainty of \DMG. Another notable groups of stars are located at around \DMG$\sim-0.75$\,mag, which are mostly the binary stars with mass-ratio close to 1. For metal-poor populations (\feh$<-0.55$), the coverage of \mpri\ is from around 0.45 to 0.7\,\Msun, while for metal-rich ones (\feh$>-0.15$), the range of \mpri\ is from 0.6 to 1.0\,\Msun.

Note that, for single stars, \mpri\ stands for the stellar mass of the single star, while for binaries, it represents for the primary stellar mass. For convenience, we do not distinguish them in notation but simply use \mpri\ for both cases. 

It is also seen that for stars with super-solar metallicity, as shown in the two bottom-right panels of Figure~\ref{fig:m1MG_feh}, the dispersion of \DMG\ is larger when \mpri$>0.9$\,\Msun. It implies that there are more stars located below the single main-sequence in the HR diagram with mass larger than 0.9\,\Msun. This larger dispersion, which could be associated with the so called blue sequence discussed by \cite{Yang2018}, induces larger uncertainties in binary fraction and mass-ratio distribution estimates. Therefore, we cut off the stars with mass larger than $0.85$\,\Msun\ to avoid it.

%%%%%%%%%%%%%%%%%%%%%%%%%%%%%%%%%%%%%%%%%%%%%%%%%%%%%%
\section{Hierarchical Bayesian model}\label{sec:method}%\label{sec:HierBayesModel}
%%%%%%%%%%%%%%%%%%%%%%%%%%%%%%%%%%%%%%%%%%%%%%%%%%%%%%

A broadly used approach to calculate the fraction of binary stars especially in star clusters is to define an empirical threshold in color-absolute magnitude diagram to separate single and binary stars. Then, a following Monte Carlo simulations, which adopt some presumed model of binaries, e.g. binary fraction and mass-ratio distribution, are conducted to compare with the observations. Subsequently, the intrinsic binary fraction and mass-ratio distribution can be obtained from the simulations which converge their results to the observations~\citep[e.g.][]{Li2013,Sana2013}.

In this work, instead to follow the previous method, we propose a hierarchical Bayesian model to simultaneously solve out the fraction of binaries and the distribution of mass-ratio of the binary systems. In Bayesian model, we concentrate to the posterior probability distribution of the binary fraction and parameters determining the shape of the mass-ratio distribution, rather than derive the best-fit binary fraction and mass-ratio distribution.

Now assume that the distribution of mass-ratio in MS binary system follows a power law, that is
\begin{equation}\label{eq:pow}
p(q|\gamma)\propto q^\gamma,	
\end{equation}
where $q$$=m_2/m_1$ represents for the mass ratio of a binary and $\gamma$ is the power index. It is noted that the power-law assumption is not a precise description of the distribution of mass-ratio~\citep{Duchene2013}. However, it is simple and can essentially provide the information about the relative number of binaries with low and high mass-ratio. 

For each MS star, the observed absolute magnitude is not only determined by its stellar mass, metallicity, and age, but also affected by two other factors: 1) the luminosity of the unresolved companion if it is a binary and 2) the random error propagated from astrometry and photometry. To quantify the two effects, we firstly need a prior knowledge about whether a star is an unresolved binary. The probability of a star to be a binary is equivalent with the fraction of binary (\fbin). If the star of interest is a binary, we also need to know in a prior about the mass-ratio of the binary so that we can derive the additional luminosity contributed by the unresolved companion. The prior knowledge about the mass-ratio is provided by the power law of the mass-ratio distribution with the power index, $\gamma$,  in Eq.~(\ref{eq:pow}). Therefore, there are two unknown parameters in the prior distributions, \fbin\ and $\gamma$. 
Then, the posterior distribution of \fbin\ and $\gamma$ given the observed differential absolute magnitude $\Delta M_{G_i}$ for the $i$th star can be written as 

\begin{align}\label{eq:bayes}
p(f_b,\gamma|&\Delta M_{i}, Z_i, m_{1,i})\propto p(f_b,\gamma|Z_i, m_{1,i})\times\nonumber\\
&\int{\sum_{c}{p(\Delta M_{i}|c,q, Z_i, m_{1,i})p(c|f_b,Z_i, m_{1,i})}\times}\nonumber\\
&\quad\quad{p(q|\gamma, Z_i, m_{1,i})dq},
\end{align}
where $Z_i$ and $m_{1,i}$ are metallicity and primary mass for the $i$th star, respectively. $\Delta M_{i}$ stands for the differential absolute magnitude. $c$ represents for the class of the star, either a single or a binary. Obviously, $p(c|f_b)$ follows Bernoulli distribution and thus the term within the integral in the right-hand side of Eq.~(\ref{eq:bayes}) can be rewritten as
\begin{align}\label{eq:mixmodel}
\int\sum_{c}&{p(\Delta M_{i}|c,q,Z_i, m_{1,i})p(c|f_b,Z_i, m_{1,i})}p(q|\gamma,Z_i, m_{1,i})dq=\nonumber\\
&(1-f_b)p_s(\Delta M_{i})\int{p(q|\gamma,Z_i, m_{1,i})dq}+\nonumber\\
&f_b\int{p_b(\Delta M_{i}|q,Z_i, m_{1,i})p(q|\gamma,Z_i, m_{1,i})dq},
\end{align}
where $p_s$ and $p_b$ stand for the likelihood distribution of $\Delta M_i$ for single and binary star, respectively. We adopt that both of them follow normal distribution and then we have

\begin{equation}\label{eq:likelisingle}
p_s(\Delta M_{i})=\frac{1}{\sqrt{2\pi}\sigma_i}\exp\left(-\frac{\Delta M_{i}^2}{2\sigma_i^2}\right)
\end{equation}
for single star and for binary it becomes
\begin{align}\label{eq:likelibinary}
p_b&(\Delta M_{i}|q,Z_i, m_{1,i})=\frac{1}{\sqrt{2\pi}\sigma_i}\nonumber\\
&\exp\left(-\frac{(\Delta M_{i}-\Delta M_{model}(q,Z_i,m_{1,i}))^2}{2\sigma_i^2}\right),
\end{align}
where $\Delta M_{model}(q,m_{1,i},Z_i)$ is the model differential absolute magnitude considering that the star has a companion with mass-ratio of $q$\ and $\sigma_i$ is the uncertainty of $\Delta M_i$.

The posterior distribution of \fbin\ and $\gamma$ for a group of stars with same (or very similar) \feh\ and \mpri\ can be obtained by multiplying all the single star posterior distributions, such as
\begin{equation}\label{eq:total_posterior}
p(f_b,\gamma|\{\Delta M_i\},Z,m_1)=\prod_{i}{p(f_b,\gamma|\Delta M_i,Z,m_1)}.
\end{equation}

\begin{figure*}
	\centering
	\includegraphics[scale=0.55]{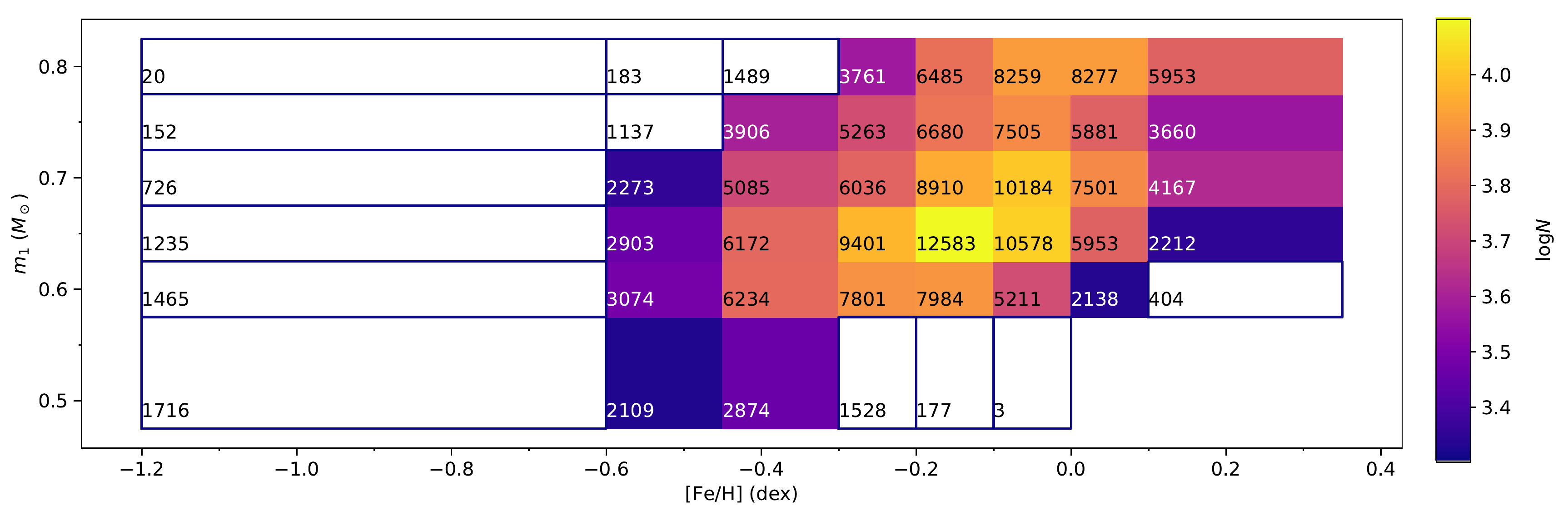}
	\caption{The figure indicates how the sample stars are separated into sub-groups in \mpri\ vs. \feh\ plane. The range of \mpri\ covers from 0.45 to 0.85\,\Msun, while the range of \feh\ covers from -1.2 to 0.4\,dex. The numbers of stars coded with colors are marked within the corresponding bins. The filled bins contain stars larger than 2000, while the blank bins contain stars less than 2000.}\label{fig:Numbers}
\end{figure*}

%%%%%%%%%%%%%%%%%%%%%%%%%%%%%%%%%%%%%%%%%%%%%%%%%%%%%%
\section{Results}\label{sec:result}
%%%%%%%%%%%%%%%%%%%%%%%%%%%%%%%%%%%%%%%%%%%%%%%%%%%%%%

We separate the stars into different metallicity and stellar mass groups, as shown in Figure~\ref{fig:Numbers}. The stars are grouped into \feh\ and \mpri\ bins with various sizes so that we can find balance between the number of stars in each bin and the resolution of metallicity and stellar mass.
Along \mpri, the center positions of the bins are at 0.525, 0.6, 0.65, 0.7, 0.75, and 0.8\,\Msun. The width of the bins are 0.15 for the first and 0.1\,\Msun\ for the other bins. The bin size of \mpri\ are a factor of 2 larger than the uncertainty of \mpri\ for individual stars. Along \feh, the centers of the bins are at -0.9, -0.525, -0.375, -0.15, -0.05, 0.1, and 0.225\,dex, with bin widths of 0.7, 0.25, 0.25, 0.2, 0.2, 0.2, 0.2, and 0.35\,dex, respectively. The bin size is also larger than the uncertainty of \feh\ for individual stars so that the errors of \feh\ would not significantly affect the results. Note that these bins are partly overlapped with their adjacent bins such that more stars are fell in the bins and therefore can stabilize the results by smoothing the results.

Table~\ref{tab:tab1} and Figure~\ref{fig:Numbers} show the numbers of stars fell into these bins. According to the Monte Carlo simulation conducted in section \ref{sec:simulation}, we find that, in principle, the results are stable when the number of stars is larger than 2000 stars in a bin. Therefore, in the rest of the analysis we only consider the bins containing larger than 2000 stars, which are filled up with colors in Figure~\ref{fig:Numbers} . This cut will lose the stars with \feh$<-0.6$.

%Figure~\ref{fig:DeltaG_m1} shows the distributions of \DMG\ for each bin with larger than 2000 stars. The selected bin size for the distribution is 0.05\,mag. It is clearly seen that the metal-poor bins (blueish lines) contains more stars with negative \DMG, i.e. binary stars, than the metal-rich populations (reddish lines). This implies that the metal-poor stars should have larger binary fraction than the metal-rich stars, as many previous works have claimed (Moe et al. 2018, Reghavan et al. 2010, Gao et al. 2015 etc). 

Then, we apply the hierarchical Bayesian model described in section~\ref{sec:method} to the \feh--\mpri\ bins with more than 2000 stars.
We run Markov chain Monte Carlo simulations using EMCEE software package~\citep{emcee}. We adopt the median values and 15\% (85\%) percentiles of the random draws in MCMCs as the best fit values and the uncertainties, respectively, for \fbin\ and $\gamma$. 

In the hierarchical Bayesian model, we set the lower limit of the mass of the secondary at 0.08\,\Msun. We do not take into account the brown dwarfs in this work.
%with \q$<0.1$ and consider them as single stars. This assumption is proper for \mpri$\sim0.8$\,\Msun\ which should not contain secondary star with mass smaller than $0.08$\,\Msun. However, for \mpri$\sim0.5$\,\Msun, this lower limit of $q$\ allows the range of mass of secondary between $0.05$ and $0.08$\,\Msun, which is unrealistic. After running the MCMC, we slightly adjust the binary fractions by removing the binaries with companions smaller than $0.08$\,\Msun. This can at most reduce 1\% to 2\% of the binary fraction. 

\begin{figure*}
	\centering	
	\includegraphics[scale=0.6]{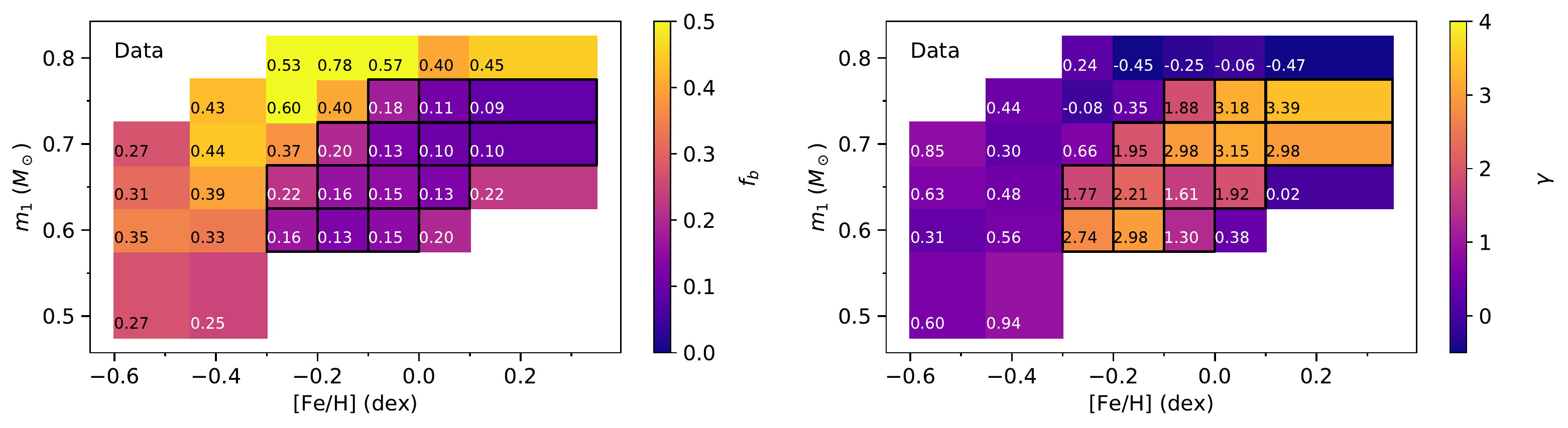}
	\caption{The left panel shows the resulting \fbin, indicated as colors, in the \mpri\ vs. \feh\ plane. The estimated \fbin\ value in each bin is marked on the plot. The right panel shows the resulting $\gamma$, indicated as colors, in the \mpri\ vs. \feh\ plane. Similar to the left panel, the estimated $\gamma$ values are also marked in the corresponding locations. The black frames in both panels indicate the bins with $\gamma>1.2$.}\label{fig:param_fehm1}
\end{figure*}

The resulting \fbin\ and $\gamma$ for different \feh\ and \mpri\ bins are shown in Figure~\ref{fig:param_fehm1} and Table~\ref{tab:tab1}. 

\begin{table*}
	\centering
	%\fontsize{6pt}{9pt}
	%\selectfont
	\caption{Best-fit \fbin\ and $\gamma$ values at different \feh\ and \mpri\ bins.}\label{tab:tab1}
	\begin{tabular}{c|ccc|ccc|ccc}
		
		\hline\hline
		\feh\ & \multicolumn{3}{c|}{$0.45<$\mpri$<0.60$}& \multicolumn{3}{c|}{$0.55<$\mpri$=0.65$}& \multicolumn{3}{c}{$0.60<$\mpri$<0.70$}\\
		\hline
		dex&N&\fbin\ & $\gamma$ & N&\fbin\ & $\gamma$ & N&\fbin\ &$\gamma$\\
		\hline
		(-0.65,-0.40)& 2109 & $0.27\pm0.03$ & $0.60\pm0.27$& 3074 & $0.35\pm0.03$ & $0.31\pm0.22$& 2903 & $0.31\pm0.03$ & $0.63\pm0.25$\\
		(-0.50,-0.25)& 2874 & $0.25\pm0.02$ & $0.94\pm0.28$& 6234 & $0.33\pm0.02$ & $0.56\pm0.16$& 6172 & $0.39\pm0.02$ & $0.48\pm0.16$\\
		(-0.35,-0.15)&&&& 7801 & $0.16\pm0.01$ & $2.74\pm0.29$& 9401 & $0.22\pm0.01$ & $1.77\pm0.21$\\
		(-0.25,-0.05)&&&& 7984 & $0.13\pm0.01$ & $2.98\pm0.42$& 12583 & $0.16\pm0.01$ & $2.21\pm0.29$\\
		(-0.15,0.05)&&&& 5211 & $0.15\pm0.02$ & $1.30\pm0.49$& 10578 & $0.15\pm0.02$ & $1.61\pm0.40$\\
		(-0.05,0.15)&&&& 2138 & $0.20\pm0.07$ & $0.38\pm0.76$& 5953 & $0.13\pm0.02$ & $1.92\pm0.59$\\
		(+0.05,0.40)&&&&&&& 2212 & $0.22\pm0.12$ & $0.02\pm0.92$\\
		\hline
		
		\feh\ & \multicolumn{3}{c|}{$0.65<$\mpri$<0.75$}& \multicolumn{3}{c|}{$0.70<$\mpri$<0.80$}& \multicolumn{3}{c}{$0.75<$\mpri$<0.85$}\\
		\hline
		dex&N&\fbin\ & $\gamma$ & N&\fbin\ & $\gamma$ & N&\fbin\ &$\gamma$\\
		\hline
		(-0.65,-0.40)& 2273 & $0.27\pm0.03$ & $0.85\pm0.32$&&&&&&\\
		(-0.50,-0.25)& 5085 & $0.44\pm0.03$ & $0.30\pm0.16$& 3906 & $0.43\pm0.04$ & $0.44\pm0.20$&&&\\
		(-0.35,-0.15)& 6036 & $0.37\pm0.03$ & $0.66\pm0.19$& 5263 & $0.60\pm0.06$ & $-0.08\pm0.17$& 3761 & $0.53\pm0.05$ & $0.24\pm0.21$\\
		(-0.25,-0.05)& 8910 & $0.20\pm0.01$ & $1.95\pm0.29$& 6680 & $0.40\pm0.04$ & $0.35\pm0.21$& 6485 & $0.78\pm0.08$ & $-0.45\pm0.15$\\
		(-0.15,0.05)& 10184 & $0.13\pm0.01$ & $2.98\pm0.48$& 7505 & $0.18\pm0.02$ & $1.88\pm0.38$& 8259 & $0.57\pm0.06$ & $-0.25\pm0.18$\\
		(-0.05,0.15)& 7501 & $0.10\pm0.01$ & $3.15\pm0.75$& 5881 & $0.11\pm0.05$ & $3.18\pm1.36$& 8277 & $0.40\pm0.05$ & $-0.06\pm0.22$\\
		(+0.05,0.40)& 4167 & $0.10\pm0.02$ & $2.98\pm1.09$& 3660 & $0.09\pm0.02$ & $3.39\pm1.46$& 5953 & $0.45\pm0.11$ & $-0.47\pm0.31$\\
		\hline\hline
	\end{tabular}
\end{table*}

%=====================================================
\subsection{Overview of the results}\label{sec:results_overview}
%=====================================================

%\subsubsection{The fractions of binary}\label{sec:result_fbin}
The left panel of Figure~\ref{fig:param_fehm1} shows that the binary fraction is roughly larger at lower \feh. Meanwhile, \fbin\ is essentially larger when \mpri\ is larger at most \feh\ values except $-0.5$\,dex. A ridge-like valley with \fbin$<0.22$ is seen from \feh$\sim-0.25$\,dex and \mpri$\sim0.6$\,\Msun\ to \feh$\sim+0.25$\,dex and \mpri$\sim0.75$\,\Msun. Within the ridge, \fbin\ increases with declining \feh. In the other regions outside this ridge, the values of \fbin\ are larger than $0.25$. Most of them are larger than $0.40$.

%\subsubsection{The power indices of the mass-ratio distribution}\label{sec:results_gamma}
The right panel of Figure~\ref{fig:param_fehm1} shows the resulting $\gamma$ as a function of \feh\ and \mpri. Similar to the distribution of \fbin, the location of the bins with $\gamma>1.2$ also concentrate in the same ridge as the low \fbin\ bins, i.e. from \feh$\sim-0.25$\,dex and \mpri$\sim0.6$\,\Msun\ to \feh$\sim+0.25$\,dex and \mpri$\sim0.75$\,\Msun. Furthermore, $\gamma$ roughly increases from $1.77$--$2.74$ at bottom-left to $3.18$--$3.39$ at top-right in the ridge. On the other hand, the values of $\gamma$ in other bins are around zero. These bins shows opposite trend to those in the ridge such that their values increase when \feh\ decreases.

\begin{figure}
	\centering
	\includegraphics[scale=0.6]{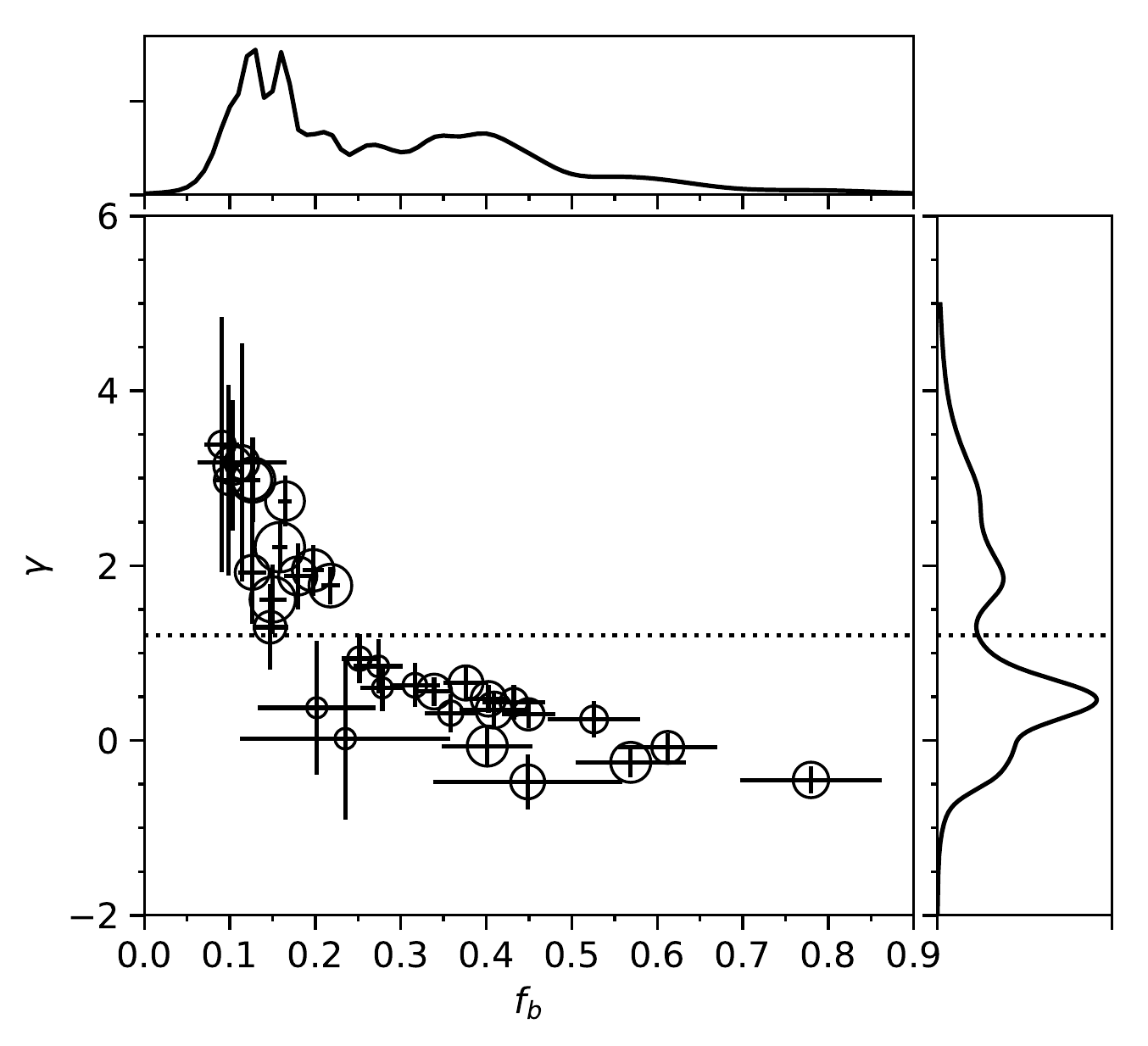}
	\caption{The figure shows the anti-correlation between the resulting \fbin\ and $\gamma$. The sizes of the circles indicate the number of samples. The larger the size the more the samples. The horizontal dotted line indicate the location of $\gamma=1.2$. The right- and top-sides show the KDE smoothed distribution of $\gamma$ and \fbin, respectively.}\label{fig:fb_gamma}
\end{figure}

Then, we draw the relationship between \fbin\ and $\gamma$ in Figure~\ref{fig:fb_gamma}. Surprisingly, we find that they are strongly anti-correlated with each other. This implies that the less the binary fraction, the more binaries with high mass-ratio and vice versa. To our knowledge, it is for the first time that the anti-correlation is unveiled in the field stars. 

The right-side panel of Figure~\ref{fig:fb_gamma} displays the distribution of $\gamma$ using KDE smoothing technique. It shows a gap at around $\gamma\sim1.2$, separating the values of $\gamma$ into two groups: the \glow\ group with $\gamma<1.2$, which contains larger binary fractions, and the \ghigh\ group with $\gamma>1.2$, which contains lower binary fractions. The \ghigh\ group is marked with black frames in Figure~\ref{fig:param_fehm1}, which is exactly overlapped with the ridge regions with low \fbin\ and high $\gamma$. 

The gap is a real feature rather than an artifact. The bin sizes are quite large than the measurement errors of \feh\ and \mpri\ and each bin contains sufficient number of samples. Therefore, the gap would not be due to the random spike dominated by noise. Moreover, the bins have been smoothed by partly overlapping with the adjacent bins. This makes the gap more robust. However, the nature of the gap is not clear.   

Since the \fbin--$\gamma$ pairs are naturally separated into two groups, in next sections, we separately discuss the features of the two groups.

\begin{figure*}
	\centering
	\includegraphics[scale=0.6]{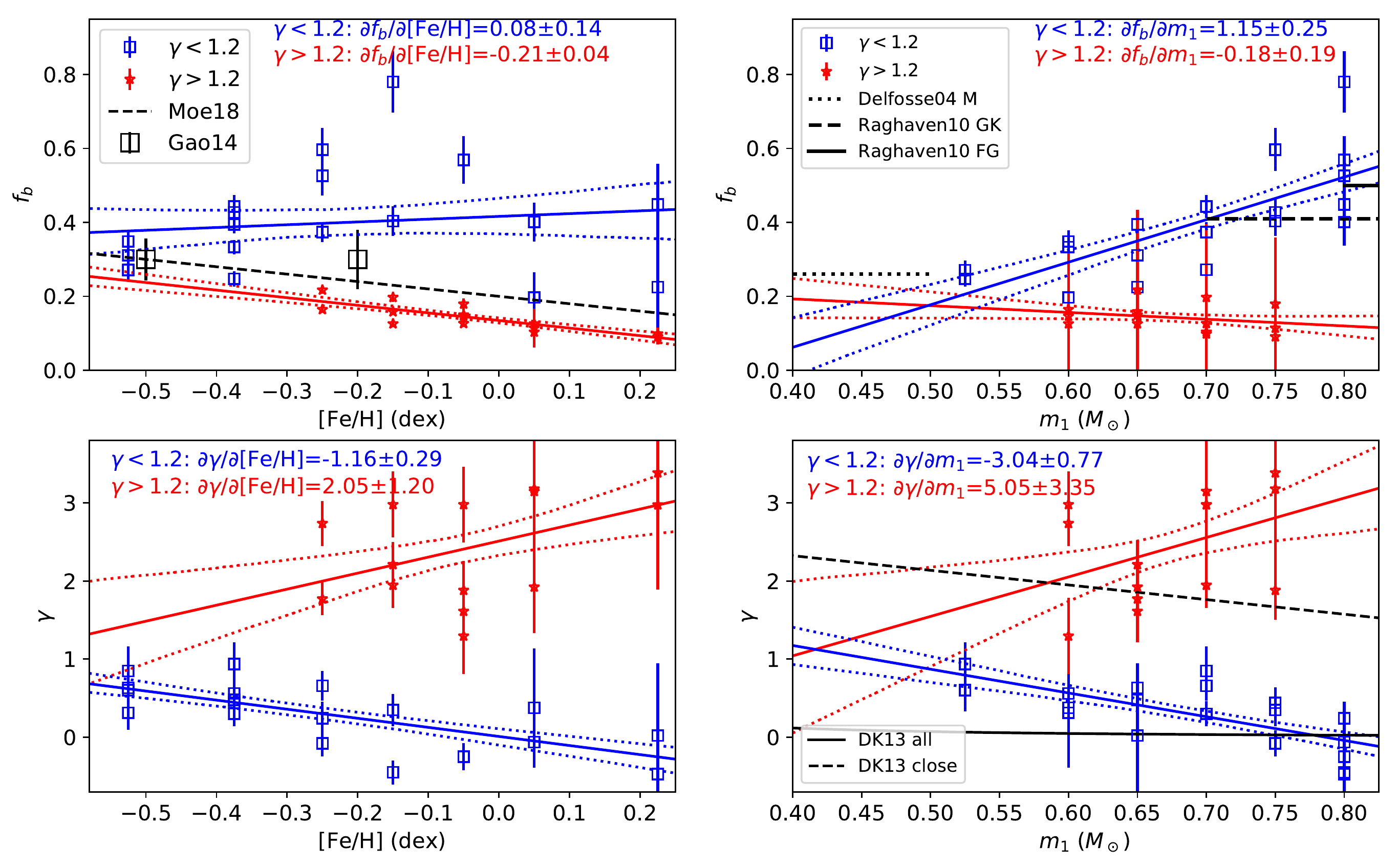}
	\caption{The results in various \feh--\mpri\ bins are separated into two groups: $\gamma<1.2$ (all in blue rectangles) and $\gamma>1.2$ (all in red stars). The top-left panel shows the distributions of the two groups in \fbin\ vs. \feh\ plane. The slopes are displayed in the panel. The result of \citet{Moe2018} is indicated as a black dashed line. The results of \citet{Gao2014} are displayed with black rectangles. The top-right, bottom-left and bottom-right panels show the distributions for the two groups in \fbin\ vs. \mpri, $\gamma$ vs. \feh, and $\gamma$ vs. \mpri\ planes, respectively. The blue and red solid lines indicate the best linear fits for groups with $\gamma<1.2$ and $>1.2$, respectively, while the dotted lines indicate the 1-$\sigma$ uncertainties of the fitting. In the top-right panel, the black dashed and solid lines indicate the $\gamma$ trend for all binaries and close binaries, respectively, mentioned in \citet{Duchene2013}. }\label{fig:gamma_groups}
\end{figure*}

%=====================================================
\subsection{Low-$\gamma$ group}\label{sec:results_gammalow}
%=====================================================
First, although the bins of the \glow\ group cover a wide range of \feh\ and \mpri, as shown in Figure~\ref{fig:param_fehm1}, only with \feh$<-0.3$\,dex does \mpri\ continuously cover from 0.5 to 0.75\,\Msun. In the regime with \feh$>-0.3$\,dex, most of the bins of the \glow\ group have \mpri\ larger than $0.7$\,\Msun. Two bins at (\feh, \mpri)=($+0.1$\,dex, $0.6$\,\Msun) and ($+0.2$\,dex, $0.65$\,\Msun) are isolated from other \glow\ bins.

Then, we look at how \fbin\ is associated with \feh\ and \mpri\ for the \glow\ group in the top panels of Figure~\ref{fig:gamma_groups}. In all four panels of the figure, the resulting \fbin\ and $\gamma$ of \glow\ group are drawn in blue. We fit the relationship of \fbin\ ($\gamma$) with \feh\ and \mpri\ respectively using a linear model. We find that \fbin\ is not related to \feh, because the slope $\partial f_b/\partial {\rm [Fe/H]}=0.08\pm0.14$\,dex$^{-1}$ (see the blue line in the top-left panel) is roughly zero. However, \fbin\ is clearly correlated with \mpri\ with substantially positive slope of $\partial f_b/\partial m_1=1.15\pm0.26$\,\Msun$^{-1}$ (see the blue line in the top-right panel). 

Finally, as shown in the bottom panels of the figure, $\gamma$ is tightly anti-correlated with both \feh\ and \mpri\ with negative slopes of $\partial\gamma/\partial {\rm [Fe/H]}=-1.16\pm0.29$\,dex$^{-1}$ and $\partial\gamma/\partial m_1=-3.04\pm0.77$\,\Msun$^{-1}$, respectively. 

%=====================================================
\subsection{High-$\gamma$ group}\label{sec:results_gammahigh}
%=====================================================

As mentioned in Section~\ref{sec:results_overview}, the \ghigh\ group bins concentrate in the ridge-like region in the \mpri\ vs. \feh\ plane (see the black frames marked in all panels of Figure~\ref{fig:param_fehm1}). This means that the \ghigh\ group only contains data with \feh$>-0.3$\,dex and \mpri\ smaller than 0.75\,\Msun. It is also noted that at the most metal-rich regions, the \mpri values of the \ghigh\ bins show lower limits, which is 0.6\,\Msun\ at \feh$=+0.1$\,dex and 0.65\,\Msun at \feh$=+0.25$\,dex.

Then, we investigate the correlation of \fbin\ with \feh\ and \mpri\ in the top panels of Figure~\ref{fig:gamma_groups}. The red stars indicate the \fbin\ values of the \ghigh\ group and the red lines display the best linear fits of the relationship of \fbin\ with \feh\ and \mpri. It shows that \fbin\ is tightly anti-correlated with \feh\ based on the substantially negative slope of $\partial f_b/\partial {\rm [Fe/H]}=-0.21\pm0.04$\,dex$^{-1}$. In the mean time, \fbin\ is not related with \mpri, since the slope $\partial f_b/\partial m_1=-0.18\pm0.19$\,\Msun$^{-1}$ is not significantly different from zero.

Finally, we look at the correlation of $\gamma$ with \feh\ and \mpri\ for the \ghigh\ group in the bottom panels of Figure~\ref{fig:gamma_groups}. We find that $\gamma$ is weakly correlated with both \feh\ and \mpri. Although the slopes $\partial\gamma/\partial {\rm [Fe/H]}=2.05\pm1.20$\,dex$^{-1}$ and $\partial\gamma/\partial m_1=5.05\pm3.35$\,\Msun$^{-1}$ are larger than zero, the significances are less than $2$-$\sigma$. The red stars located in the bottom panels also show larger dispersion than the \glow\ group (blue rectangles). Therefore, we infer that the correlations of $\gamma$ with either \feh\ or \mpri\ is quite marginal.

To summarize these complicated relationships, we list them in Table~\ref{tab:sum_relat}.

\begin{table}
	\caption{Summary of the relationships in $\gamma$-low and high groups.}\label{tab:sum_relat}
	\centering
	\begin{tabular}{c|c|c}
		\hline\hline
		Condition & $\gamma<1.2$ & $\gamma>1.2$\\
		\hline
		\fbin\ vs. \feh\ & no relation & anti-correlation\\
		\fbin\ vs. \mpri\ & correlation & no relation\\
		$\gamma$ vs. \feh\ & anti-correlation & marginal correlation\\
		$\gamma$ vs. \mpri\ & anti-correlation & marginal correlation\\
		\hline
		\mpri\ & larger \mpri\ & smaller \mpri\\
		\feh\ & lower \feh\ & higher \feh\\ 
		\hline\hline
	\end{tabular}
\end{table}

%=====================================================
\subsection{Smoking gun of the dynamical processing}\label{sec:result_explain}
%=====================================================
In this section, we explain the results described in above sections based on the current theories of binary formation and evolution. 
We firstly  give four assumptions.

First, we assume that the mass function of the secondary is universal, i.e. no matter \feh\ and \mpri, the secondary mass function are all the same. It is not necessarily to be the canonical IMF given by \citet{Chabrier2005} or \citet{Kroupa2003} or others. Note that although \citet{Myers2011} has discussed theoretically about the invariability of the IMF with metallicity, it is not clear whether the IMF of secondary is same as that of the primary.%This assumption is quite trivial such that no matter the binary is formed from gas or disk fragmentation the mass function of the companion is quite natural to follow the same rule as in the single star.

%While, early literatures prefer to the core fragmentation as the dominate channel to form binaries \citep{Tohline2002}, more and more observational evidence support that the disk fragmentation may be the prominent way~\citep{Moe2018}.
Second, according to \citet{Moe2018}, the close binaries are more likely formed from disk fragmentation, which is associated with the metallicity of the gas~\citep[also c.f.][]{Kratter2008,Kratter2010,Machida2009,Tanaka2014}. However, for the wide binaries, as found by \citet{ElBadry2019}, they are not related to metallicity, which is consistent with the scenario that the wide binaries are formed from gas fragmentation~\citep{Tohline2002,Offner2010,Myers2011,Bate2012,Bate2014}. Consequently, we assume that the binaries are formed both from  disk fragmentation, which contributes mostly to the close binaries depending on metallicity, and gas fragmentation, which produces most of the wide binaries not depending on metallicity.

Third, we assume that the binary fraction increases with primary stellar mass based on the simulation by \citet{Tanaka2014}. This assumption is also supported by many observations~\citep{Delfosse2004,Raghaven2010,Sana2012,Sana2013}.

Finally, it is known that most stars are formed in embedded clusters~\citep{Lada2003}. \citet{Marks2011} argued that the dynamical processing in the embedded clusters, in which binaries were born and spent their early times, prefers to efficiently disrupting binaries with larger separations, smaller primary mass, and smaller mass-ratios. This is because the binding energy of a binary system can be expressed as
\begin{equation}\label{eq:bindE}
E_b \propto \frac{m_1^2q}{a},
\end{equation}
where $a$ is the separation of the two companions~\citep*{Marks2011a}. The dynamical processing occurred in embedded clusters has a very short time scale of a few million years~\citep{Belloni2018}. We assume that the dynamical processing is generally very efficient and can change the present-day field binary properties to different extents depending on the primary mass, separation, mass-ratio etc.

Now, we can apply all above assumptions to the two groups of stars and explain why they show the observed \fbin\ and mass-ratio distribution features as listed in Table~\ref{tab:sum_relat}.

Firstly, because the \ghigh\ group contains smaller \mpri, they are easier to be affected by the dynamical evolution. The dynamical processing destroys many binaries with lower $q$ and wider separation since they have smaller binding energy. Consequently, this leads to smaller binary fraction for \ghigh\ populations.
In the mean time, the loss of lower mass-ratio binaries also results in larger $\gamma$ since the existing binaries are more biased to larger $q$.
Moreover, the loss of wide binaries tends to increase the fraction of close binaries in the present-day samples. According to the second assumption that the formation of close binaries is metallicity-dependent, the close binary dominated populations exhibit the tight anti-correlation between \feh\ and \fbin.

%There are a few evidences to support this explanation. First, the \ghigh\ group has a significantly smaller binary fraction because of the efficient disruption. Second, the dynamical disruption favorites wide binaries and leave more close binaries. This leads to the \fbin\ of the \ghigh\ group is anti-correlated with \feh\ according to the second assumption.
%This may lead to the reshaping of the period distribution~\citep{Marks2011}. Note that the binaries in our data is mostly discriminated from the photometry, which does not reflect the period or the separation of a binary. However, we can still indirectly check this effect by looking at the metallicity. 
%The close binaries may become more prominent in the surviving binaries since the wide binaries are more effectively disrupted. According to the second assumption, the close binaries are likely formed from disk fragmentation and thus the close binary fraction is anti-correlated with metallicity. This leads to a substantial anti-correlation between \fbin\ and \feh\ in the \ghigh\ group. 
%Finally, the dynamical processing prefer to destroying binaries with lower primary mass because of their lower binding energy. This can explain why the \ghigh\ group bins are dominated by low primary mass stars. The \ghigh\ group also shows lower limit of \mpri. This may be due to that the very low mass stars can be effectively ejected from the embedded clusters before they are affected by the dynamical process within the clusters.

Meanwhile, the \ghigh\ group is not only biased to lower \mpri, but also  biased to higher metallicity. This means that the metal-rich stars may have experienced more effective dynamical processing. It is not clear how metallicity plays the role in the formation of the embedded clusters and how it affect the dynamical processing. This observational evident may be helpful in clarifying how metallicity works in this issue. 

On the other hand, because the \glow\ group contains larger primary masses, it is less affected by the dynamical processing. Therefore, the populations in this group have relatively larger present-day binary fractions. The flat \fbin--\feh\ relationship is evident that the populations with smaller $\gamma$ contain more metallicity-independent wide binaries, which smear out the \fbin--\feh\ anti-correlation contributed by close binaries. These support that the \glow\ group may be less or even not affected by the dynamical processing during the stage of the embedded clusters.

The anti-correlations between $\gamma$ and \feh\ and between $\gamma$ and \mpri\ in \glow\ group may reflect the various star (and also binary) formations influenced by various characteristics, e.g. cluster mass, IMF, metallicity, of the embedded clusters. Future theoretical modeling making use of these observational evidences will be very useful to investigate the nature of star formation.

We should emphasize that the explanations of high- and \glow\ groups are based on a number of assumptions compiled from various theoretical and observational works. Because the theoretical works are usually limited by their ingredients and techniques, they do not exactly reflect the realities. Therefore, these explanations are not exclusive. Although it seems that they are well consistent with the assumption and the observations, other mechanisms are not ruled out.

%=====================================================
\subsection{Identification of binaries}\label{sec:result_binarycand}
%=====================================================

\begin{figure}
	\centering
	\includegraphics[scale=0.7]{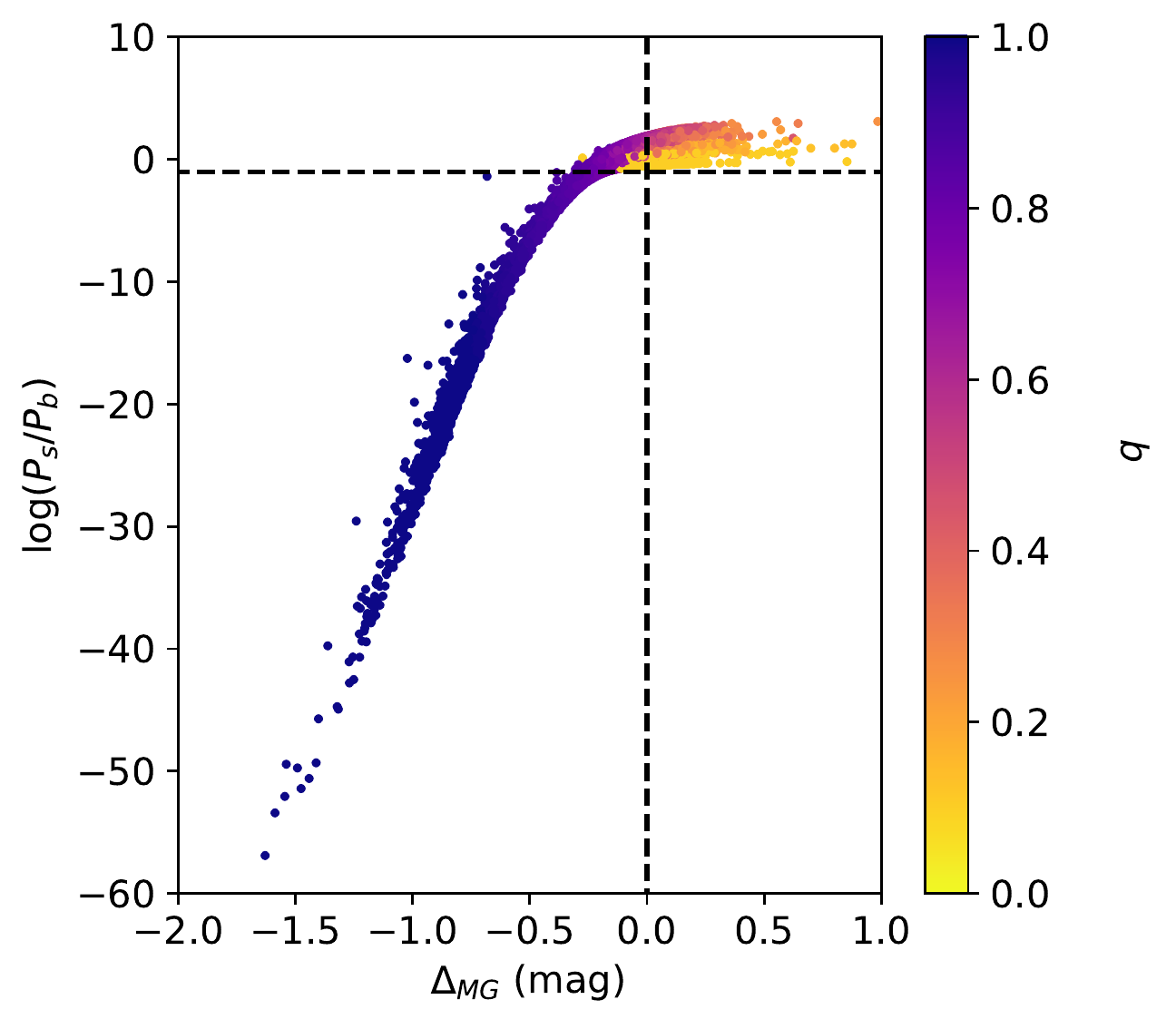}
	\caption{This displays the relation between $\log P_s/P_b$ and \DMG. The horizontal dashed line indicates $P_s/P_b=0.1$. The colors code the mass-ratio if a star is a binary.}\label{fig:binary_odd}	
\end{figure}

In each \feh--\mpri\ bin, we are able to identify which star is likely a binary, since the resulting \fbin\ and $\gamma$ are the hyper-parameters of the prior in Eq.~(\ref{eq:bayes}). We can then use Eq.~(\ref{eq:mixmodel}) to estimate the probabilities that a star to be either a single or a binary. Specifically, the probability of the $i$th star to be a single can be written as
\begin{equation}
P_{s,i} = (1-f_b)p_s(\Delta M_i)\int{p(q|\gamma)dq}.
\end{equation} 
And the probability to be a binary reads
\begin{equation}
P_{b,i} = f_b\int{p_b(\Delta M|q)p(q|\gamma)dq}.
\end{equation}
Then, the odd of the star to be a single can be written as $P_{s}/P_{b}$. When $P_{s}/P_{b}$ is smaller than 1, the star is more likely to be a binary, while it is larger than 1, it is likely to be a single star. Figure~\ref{fig:binary_odd} shows that $\log P_s/P_b$ is correlated with \DMG\ as expected. Given a star, the more likely to be a binary, the smaller the \DMG.
%We adopt the separation of single and binary stars at  $P_{s}/P_{b}=0.1$. Figure~\ref{fig:binary_odd} shows the relationship between $\log P_s/P_b$ and \DMG. It displays that, for a star, the smaller the \DMG, the more likely to be a binary. 

When a star is identified as a binary, the mass-ratio can also be roughly approximated from \DMG. We consider a posterior distribution for $q$\ such that
\begin{equation}\label{eq:massratio}
p(q|\Delta M)=p(\Delta M|q)p(q|\gamma).
\end{equation}
We adopt the best-fit $q$\ to be at the maximum of the posterior distribution and the 15\% and 85\% percentiles of the distribution as the uncertainties. Figure~\ref{fig:binary_odd} also indicates the mass-ratios based on Eq.~(\ref{eq:massratio}).

%of the classified binaries with colors. It is noted that only when \q$\gtrsim0.7$, the odd can robustly favorite a binary, while the low mass-ratio binaries are likely classified as single stars.
%Although it is difficult to individually identified the binaries with \q$<0.7$, we are able to derive the fraction of binary and the slope of the mass-ratio distribution given the analytic form of mass-ratio distribution in the hierarchical Bayesian model. 

Because only the filled regions in Figure~\ref{fig:Numbers} have resulting \fbin\ and $\gamma$, only the stars located in these regions can be classified as single and binaries. 
We finally calculate $\log P_s/P_b$ and $q$ for 57\,978 stars with \feh$\gtrsim-0.6$\,dex. 
Table~\ref{tab:star} lists the whole 57\,978 samples with $\log P_s/P_b$ and the best-fit mass-ratio. Note that $q$ estimated from Eq.~(\ref{eq:massratio}) is only related to \MG\ and $\gamma$, no matter whether a star is classified as binary or single, we provide $q$ value and its uncertainty for it. We suggest 5\,410 of these samples with $\log P_s/P_b<0.1$ are very likely binaries.
\begin{table*}
	\centering
	\caption{The MS star catalogue with probability of binary and mass-ratio. The full catalogue is at \url{https://github.com/liuchaonaoc/Binarity/blob/master/table3.txt}.}\label{tab:star}
	\fontsize{6pt}{9pt}
	\selectfont
	\begin{tabular}{c|c|c|c|c|c|c|c|c|c|c|c|c|c}
		\hline\hline
		LM obsid & \gaia\ source id & ra & dec  & $M_G$ & $M_G$ error & $A_G$ & $\Delta M_G$ & $m_1$ & $m_1$ error & $\log P_s/P_b$ & $q$ & $q$ 15\% & $q$ 85\%\\ 
		& & h:m:s & d:m:s & mag & mag & mag & mag & \Msun\ & \Msun\ & & & & \\
		\hline
		601119 & 1893391222402009728 & 22:11:51.26 & +28:37:44.8 & 7.217 & 0.083 & 0.313 &  0.067 & 0.57 & 0.03 & 1.96 & 0.46 & 0.29 & 0.56\\
		603081 & 1896846231534740352 & 22:04:18.06 & 	+30:22:39.0 & 5.866 & 0.082 & 0.261 & -0.962 &0.65&0.04& -26.62& 1.0	& 0.98 & 1.0\\
		603241 & 1896831594286160384 & 22:04:38.07	& +30:12:40.5 & 6.282 & 0.083 & 0.251 & 0.072 &0.71&0.04& 1.4	& 0.51	& 0.29 & 0.61\\
		  604003  &  1895407486209553664  & 22:10:48.25 &   +30:45:29.4  & 6.692 &  0.081 & 0.236    &   -0.115  &  0.68 &   0.03  &     1.04    &      0.72 &  0.54 &   0.76\\
		\hline\hline\\
	\end{tabular}
\end{table*}

%%%%%%%%%%%%%%%%%%%%%%%%%%%%%%%%%%%%%%%%%%%%%%%%%%%%%%
\section{Discussions}\label{sec:disc}
%%%%%%%%%%%%%%%%%%%%%%%%%%%%%%%%%%%%%%%%%%%%%%%%%%%%%%

%=====================================================
\subsection{Comparison with other works}\label{sec:disc_compare}
%=====================================================

%=====================================================
\subsubsection{Binary fraction}\label{sec:fbin}
%=====================================================
We compare the binary fraction from our results with \citet{Moe2018} (black dashed line in the top-left panel of Figure~\ref{fig:gamma_groups}). The slope of the anti-correlation between \fbin\ and \feh\ is $\partial f_b/\partial{\rm[Fe/H]}\sim-0.12$\,dex$^{-1}$ from \citet{Moe2018}, which is similar to $-0.21\pm0.04$\,dex$^{-1}$ from the \ghigh\ group. The binary fraction from \citet{Moe2018} is larger by about 0.1 than that of the \ghigh\ group. Compared with the binary fraction of \glow\ group, however, we find that the \fbin\ from \citet{Moe2018} is significantly smaller. Because \citet{Moe2018} studied the close binaries ($a<10$\,AU), it is not surprising that their \fbin\ is more similar to \ghigh\ group, which is likely affected by dynamical evolution and hence more dominated by close binaries, than to \glow\ group, which contains more wide binaries.
%
%In the mean time, we compare the anti-correlation between \fbin\ and \feh\ with \citet{Moe2018} using the slope of the linear relationship of \fbin\ with \feh. For the \glow\ group, the slope $\partial f_b/\partial{\rm[Fe/H]}$ is $-0.21\pm0.04$\,dex$^{-1}$ and it is $-0.12$\,dex$^{-1}$ derived from the results in \citet{Moe2018}, which are consistent with each other.

%However, the \fbin\ of \ghigh\ group is larger and its $\partial f_b/\partial{\rm[Fe/H]}$ is flatter than \citet{Moe2018}. This is probably because that \ghigh\ group contains more binaries with long period. 

\citet{ElBadry2019} shows that the anti-correlation is substantial in binaries with separation smaller than 100\,AU but disappears in wider binaries. This also agrees with our results for \ghigh\ and -low groups.

We compare the \fbin--\mpri\ correlation with previous works. First, \citet{Delfosse2004} and \citet{Leinert1997} found that the multiplicity fraction of M dwarf stars ($0.1$--$0.5$\,\Msun) is $0.26$. The closest mass range in our work is \mpri$\sim0.525$\,\Msun, which has \fbin\ of $0.25\pm0.02$. This is consistent with \citet{Delfosse2004}. %no matter gamma
Second, \citet{Raghaven2010} selected stars in the mass range slightly larger than us. 
%Subsequently, their binaries may less affected by the dynamical evolution due to the larger \mpri. Therefore, they should be comparable with the \glow\ group.  
The multiplicity fraction of F6-G2 stars, which are slightly more massive than $\sim0.8$\,\Msun,  in \citet{Raghaven2010} is $0.50\pm0.04$. 
Only \glow\ group contains \mpri$\sim0.8$\,\Msun and it falls between $0.40$ and $0.78$ depending on different metallicity. The fraction of their G2-K3 samples, which covers $0.7<$\mpri$<1$\,\Msun, is $0.41\pm0.03$, while our results with \mpri$=0.7$\,\Msun\ at \glow\ group is from $0.40$ to $0.60$, which are in the similar ranges.

%=====================================================
\subsubsection{Mass-ratio Distribution}\label{sec:massratio}
%=====================================================

We then compare the resulting $\gamma$ with previous literatures. \citet{Duquennoy1991} argued that the mass-ratio distribution of the G type stars is quite similar to the canonical IMF \citep*{Miller1979,Kroupa1990}, while \citet{Raghaven2010} suggested a flat distribution of mass-ratio with a sharp peak at $q\sim1$. However, we do not find any literature has discussed about the anti-correlation of the mass-ratio distribution with metallicity as displayed in the bottom-left panel of Figure~\ref{fig:gamma_groups} for the \glow\ group.

\citet{Duchene2013} compiled many literatures and demonstrated that the power index of a power-law fitted mass-ratio distribution substantially changes with the primary mass (see their Figure 2). At \mpri$\lesssim1$\,\Msun, the smaller the primary mass, the larger the power index. This means that for low-mass binaries, the secondaries tend to have similar mass as the primaries, while the more massive binaries tend to have more low-mass companions. We fit the $\gamma$ values for binaries with all range of period and \mpri$<1$\,\Msun, which are displayed as the red diamond symbols in the Figure 2 of \citet{Duchene2013}, using a power law and show it as black solid line in the bottom-right panel of Figure~\ref{fig:gamma_groups}. In the range of \mpri\ between 0.4 and 0.85\,\Msun, the $\gamma$s derived from previous works prefer to a quite flat relationship with \mpri. Compared to the \glow\ group, at around \mpri$\sim0.8$\,\Msun, our result is similar to \citet{Duchene2013}. However, the $\gamma$ of the \glow\ group has stronger anti-correlation with \mpri\ than \citet{Duchene2013}. \citet{Duchene2013} also provided $\gamma$ for populations of binaries with smaller period (the blue squares in their figure 2). We simply connect the two points at \mpri$\sim0.2$ and $1.0$\,\Msun\ and show it with a black dashed line in the bottom-right panel of Figure~\ref{fig:gamma_groups}. Their $\gamma$ is essentially similar to the values of the \ghigh\ group. This is also consistent with the hypothesis that \ghigh\ group is dominated by shorter period binaries due to the effective dynamical processing.

\subsection{Effect of sample size}\label{sec:simulation}
%=====================================================

\begin{figure*}
	\begin{minipage}{18cm}
		\centering
		\includegraphics[scale=0.5]{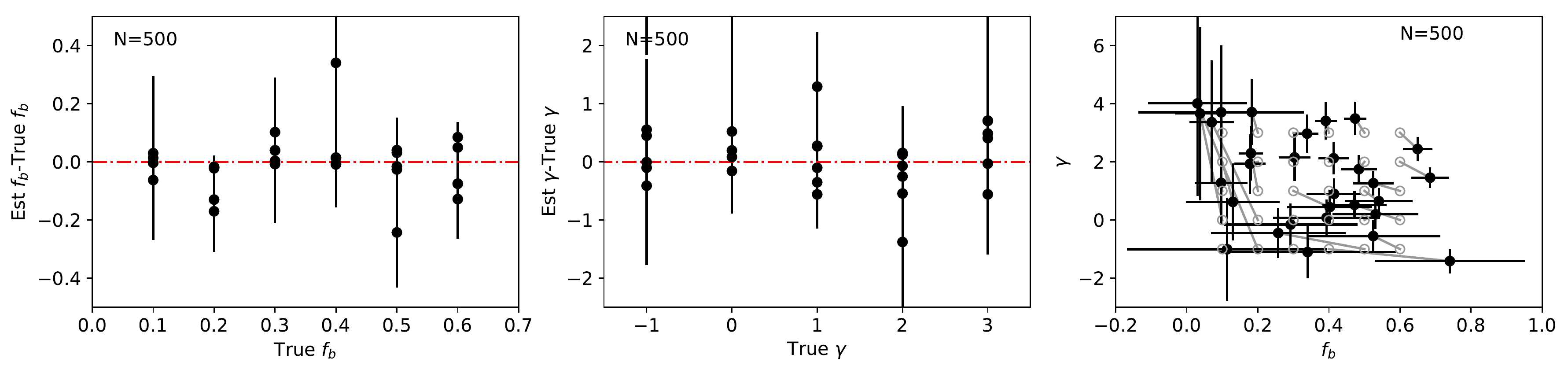}
		\includegraphics[scale=0.5]{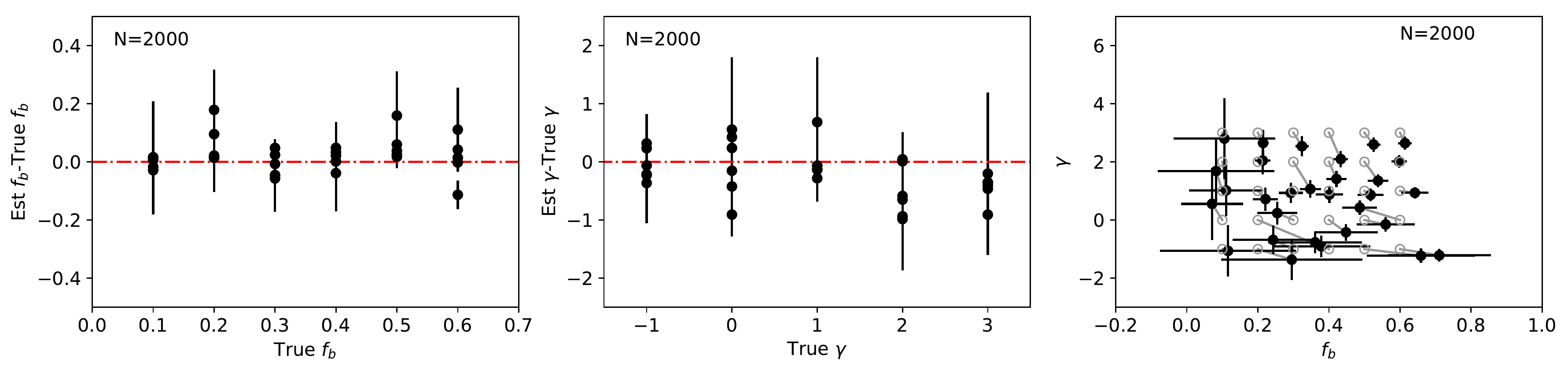}
		\includegraphics[scale=0.5]{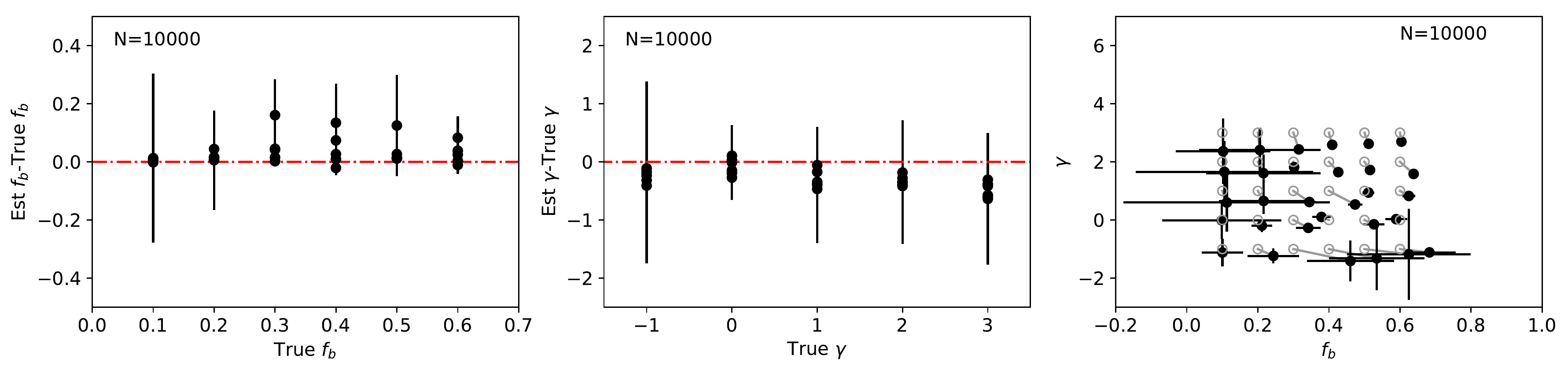}
	\end{minipage}
	\caption{The rows show the results  of simulations to validate the hierarchical Bayesian model with the number of mock stars of 500, 2000, and 10000, from top to bottom, respectively. The left column shows the difference of the estimated \fbin\ with the true ones. The middle columns shows the difference of $\gamma$ estimates with the true values. The right column shows the difference of the estimated pairs of \fbin\ and $\gamma$ (black dots with error bars) compared with the true pairs (gray hollow circles).}\label{fig:simulation}
\end{figure*}
%In many studies about the binaries in the open clusters, only a few hundreds of stars are involved to derive the binary fraction. Few works have investigated that with how many stars the binary fraction estimates is robust. 
In this section, we investigate the effect of the sample size in the hierarchical Bayesian model.

For this purpose, we generate mock samples with various numbers of mock stars and different \fbin\ and $\gamma$. For all simulated data, we adopt the primary stellar mass \mpri$=0.7$\,\Msun\ and \feh$=-0.3$\,dex. We first select 6 \fbin\ values at $0.1$, $0.2$, ..., $0.6$ and at 5 $\gamma$ values at $-1$, $0$, ..., $3$. Given any combination of \fbin\ and $\gamma$ from the above values, we arbitrarily draw three mock datasets with numbers of $N=500$, $2000$, and $10000$, respectively. Each mock star has a probability of \fbin\ to be a binary. If it is a binary, the mass-ratio $q$ is arbitrarily drawn from the power law with the given $\gamma$. In total, we have $6\times5\times3=150$ groups of mock data with different \fbin, $\gamma$, and samples size. 

The differential absolute magnitude $\Delta M_G$ for the mock data is calculated from PARSEC stellar model given the adopted \mpri, \feh, and $q$. We then run the hierarchical Bayesian model described in section~\ref{sec:method} via MCMC for each group of the mock data.

Figure~\ref{fig:simulation} shows the results of the mock data grouped with sample size. The first row shows the results for the mock data with $N=500$. The top-left panel displays that the estimated \fbin\ is essentially consistent with the \emph{true} values (the presetting binary fraction when generating the mock data, $x$-axis). However, in the top-middle panel, a few estimated $\gamma$ are substantially biased, while the others well reproduce the ``true'' $\gamma$. The top-right panel shows the offsets of the estimated (black dots with error bars) parameters from the presetting parameters (gray circles). It is seen that when the true \fbin\ is smaller than $0.25$, the $\gamma$ tends to be overestimated. Meanwhile, the binary fractions are slightly overestimated with larger error bars to around $0.2$ when the presetting fraction is larger than $0.25$ and $\gamma<1$. The overall systematic offset of derived \fbin\ is $-0.01$ with uncertainty of $0.10$. The systematic bias of the estimated $\gamma$ is $0.46$ with uncertainty of $1.38$ for the sample size of $N=500$.

The middle row of Figure~\ref{fig:simulation} shows the results for the mock data with $N=2000$. Because the sample sizes increases, the performance of the estimations of \fbin\ and $\gamma$ are substantially improved. The left and middle columns show that the estimates of \fbin\ and $\gamma$ well reproduce the presetting values. The right column shows no significant overestimation for $\gamma$ at \fbin$<0.2$, while the slightly overestimation and larger uncertainty of \fbin\ still exists. The $\gamma$ is also slightly underestimated in most of the datasets at \fbin$>0.2$ and $\gamma<0$. At \fbin$<0.2$, the uncertainties of both $\gamma$ and \fbin\ are large. The systematic bias of the estimated \fbin\ is $0.02$ with uncertainty of $0.06$. The bias of the deriving $\gamma$ is $-0.21$ with dispersion of $0.47$. 
%When $\gamma\leq0$, the systematic bias of \fbin\ increases to $\sim0.04$. 
The random errors are significantly smaller than the results of the $N=500$ samples and the systematic bias in $\gamma$ is reduced by a factor of 2.

The bottom row shows the results for samples with $N=10000$. The bias of the \fbin\ estimates is $0.03$ with dispersion of $0.05$. For $\gamma$ estimates, they are $-0.29$ and $0.34$. Compared to the results for $N=2000$, the systematics does not improve any more, although the random errors mildly improve. 

To summarize, the exercises with mock stars are evident that when the sample size is a few hundreds, the binary fraction estimates is reliable but suffers from large random errors. The mass-ratio distribution estimates, however, may be significantly overestimated when binary fraction is low. When $N\gtrsim2000$, the estimations for \fbin\ and $\gamma$ are already robust. Although increasing sample size can reduce the random error, a slight systematic bias does not disappear.

We therefore conclude that, first, we need to keep the sample size to be larger than 2000 so that the random errors and systematics can be well controlled. This is why we only derive \fbin\ and $\gamma$ for the \feh\--\mpri\ bins with $N>2000$. Second, the readers have to be aware of that the results shown in section~\ref{sec:result} may slightly overestimate the binary fraction by around $0.03$ and underestimate $\gamma$ by $0.2$--$0.3$. Finally, although the covariance between \fbin\ and $\gamma$ does exist and leads to larger uncertainties when either \fbin\ or $\gamma$ is small, there is no strong degeneracy between \fbin\ and $\gamma$, meaning that they can be determined with reasonable uncertainties and thus, the anti-correlation shown in Figure~\ref{fig:fb_gamma} is real.

%=====================================================
\subsection{More discussion about the \fbin--$\gamma$ anti-correlation}\label{sec:fb_gamma_anti}
\begin{figure}
	\centering
	\includegraphics[scale=0.6]{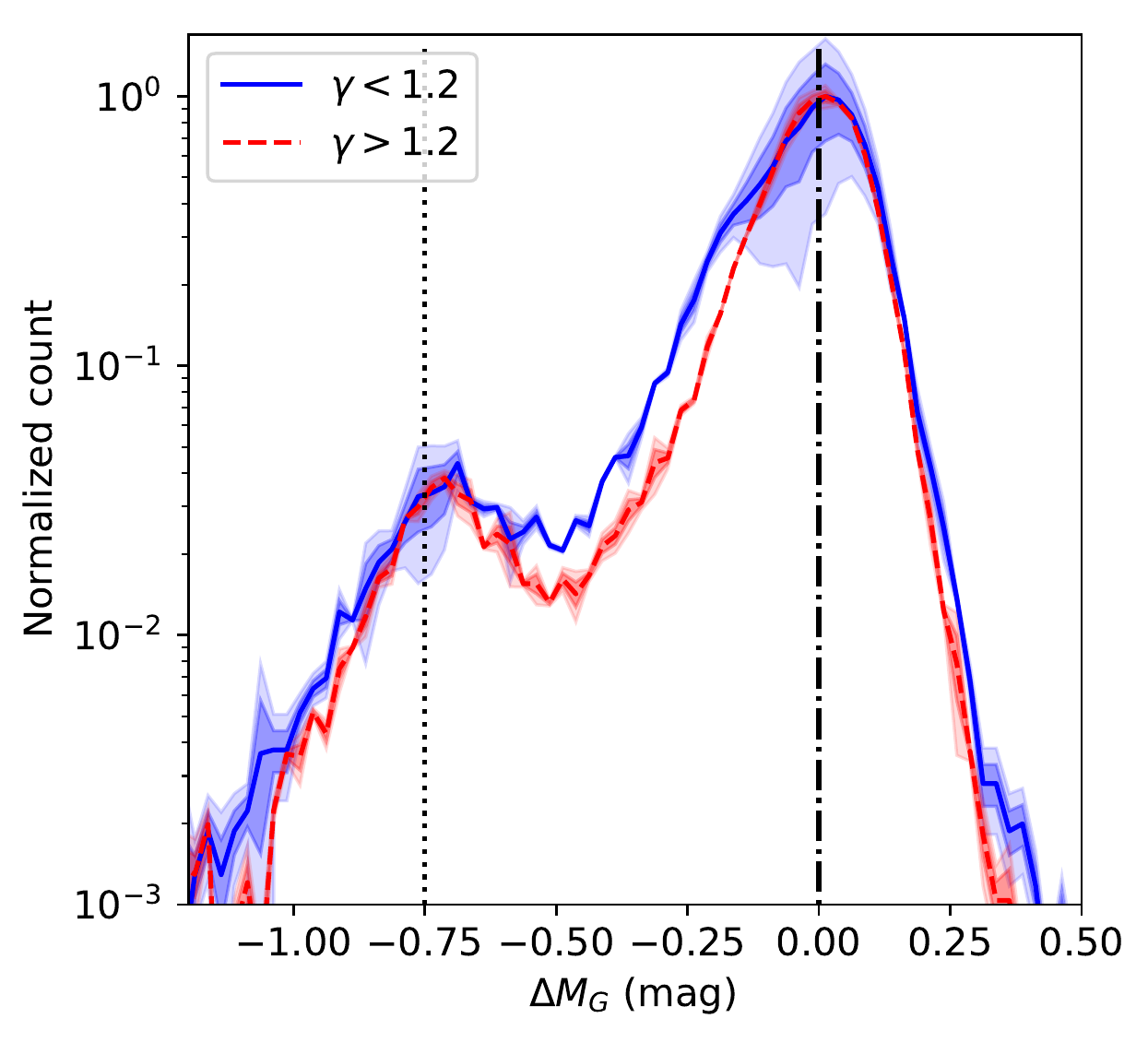}
	\caption{Two distributions of \DMG\ for the data with $\gamma>1.2$ (blue line) and $\gamma<1.2$ (red). The shadows show $1-$ and $2-\sigma$ uncertainties of the distributions. $y-$axis is displayed in logarithmic scale. The vertical black dot-dashed line indicate the location of single stars and the black dotted line indicate the binaries with $q=1$.}\label{fig:DMGsample}	
\end{figure}

%=====================================================
To double-check the anti-correlation between \fbin\ and $\gamma$, we compare the mean distributions of \DMG\ for the two groups of stars in Figure~\ref{fig:DMGsample}. The two distributions are normalized so that the maximum values are at 1. The \glow\ group (blue line) shows more fraction of stars at \DMG\ between $-0.5$ and $-0.1$\,mag than the \ghigh\ group, while the two groups have similar fraction of binaries at \DMG$\sim-0.75$\,mag. This means that the \glow\ group contains more binaries with moderate $q$ than the \ghigh\ group, although they have similar fractions of binaries with $q$$\sim1$. This leads to a two-fold affect. First, the fraction of binaries for the \glow\ group is larger, since it contains more binaries with moderate $q$. Second, $\gamma$ must be smaller for the \glow\ group, because that its mass-ratio distribution must be flatter than the \ghigh\ group since it contains more binaries with moderate $q$. This again confirms that the anti-correlation between \fbin\ and $\gamma$ is indeed a real feature.

%=====================================================
\subsection{Incompleteness of period distribution}\label{sec:period}
%=====================================================

\begin{figure}
	\centering
	\includegraphics[scale=0.55]{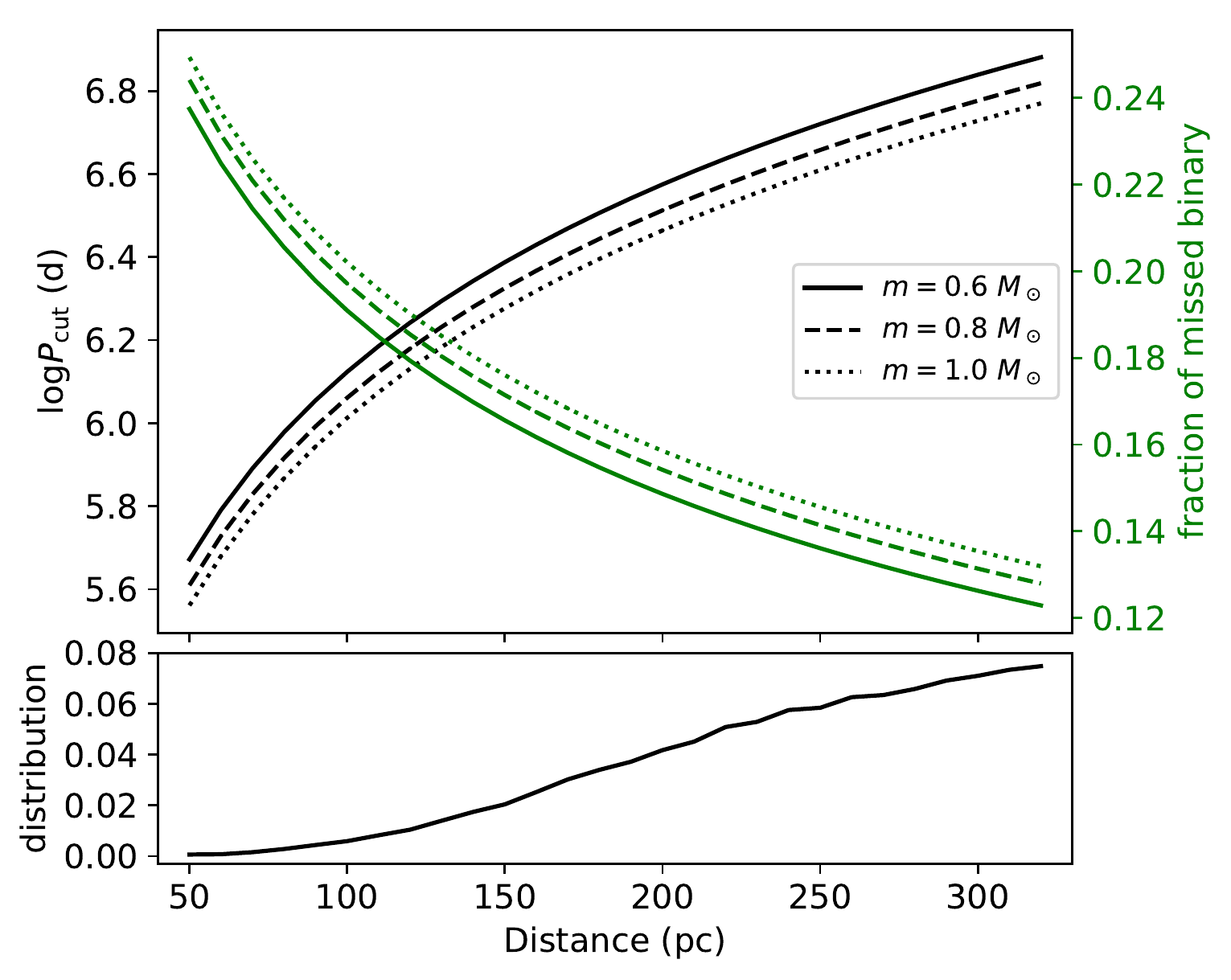}
	\caption{The black lines aligned with the left-hand side $x$-axis shows the logarithmic cut-off period subject to the 2\,arcsec angular resolution of \emph{Gaia} as a function of distance. The green lines aligned with the right-hand side $x$-axis displays the fraction of the missing binaries due to the period cut-off. The solid, dashed, and dotted lines indicate three cases with total masses of $0.6$, $0.8$, and $1.0$\,\Msun, respectively. The bottom panel indicates the number distribution of the samples along the distance.}\label{fig:missingbin}	
\end{figure}

In this work, we identify unresolved binaries from HR diagram, which requires the binaries to appear as a single point. When a binary shows recognizable spatial separation between the two companions, it would be mistakenly identified as two single stars rather than one binary. Therefore, our method may miss wide binaries in the samples and lead to incompleteness to some extent. %In general, a binary located at nearer distance to the Sun is more likely resolved as two single stars. This means that the fraction of missing wide binaries in our samples varies with distance. 
In this section we evaluate the effect of the incompleteness.

The incompleteness of period distribution is mainly subject to the angular resolution of \emph{Gaia},which is around $2$\,arcsec \citep{Arenou2018}. The corresponding cut-off period varies with distance, as shown with black lines in Figure~\ref{fig:missingbin}. The cut-off in period leads to incompleteness of binaries in distribution of period. We then approximate the fraction of missing binaries with large periods beyond the cut-off points by adopting the distribution of period from \citet{Raghaven2010}.  The fractions of missing binaries is displayed as a function of distance with green lines in Figure~\ref{fig:missingbin}. It is seen that, in the worst case at 50\,pc, about one fourth binaries are missed. However, at 300\,pc, only 13\% binaries, most of which have wide separations, are cut off. Weighted by the number distribution of the observed stars along distance (bottom panel in Figure~\ref{fig:missingbin}), the averaged fraction of the missing binaries is around 15\%. 

The missing wide binaries with period beyond the cut-off points become two single stars to the field. The secondaries of the missing binaries may have opportunity to be observed by LAMOST as an independent single star. However, since the secondary is fainter, they have less probability to be targeted in LAMOST survey according to the selection function of LAMOST, which is roughly flat along apparent magnitude~\citep{Carlin2012, Yuan2015, Liu2017}. Moreover, lots of the resolved secondaries have masses lower than 0.45\,\Msun, which have been already excluded from our samples. 
We can reasonably assume that the completeness of LAMOST survey is around 20\%, i.e., only one fifth of the stars in a field are finally observed by LAMOST. And we can coarsely assume that about half of the resolved secondaries observed by LAMOST have masses larger than 0.45\,\Msun\ and hence are considered as single stars. Combining these two factors, we obtain that only 10\% of the resolved secondaries from the 15\% missing wide binaries mistakenly contribute to the single stars. This means that the number of the single stars only increase by a few percents due to the incompleteness of the wide binary.
Therefore, we ignore the systematic bias of the number of single stars in our samples. 

Finally, considering that the resulting binary fraction is slightly overestimated (see section~\ref{sec:simulation}) by $\sim0.02$, the systematic underestimation of the binary fraction due to the incompleteness of the distribution of period reduces from 15\% to 13\%.

%=====================================================
\subsection{Effect from high order multiple systems}\label{sec:multiplicity}
%=====================================================
Because we identify binaries from their larger luminosity, we cannot distinguish the binary and high-order multiple system. Therefore, \fbin\ used in the whole paper is actually equivalent with the frequency of multiplicity, i.e. how many observed stellar objects are multiple stars. This is different with the component frequency, which is defined as the averaged number of companions in each stellar object \citep{Duchene2013}.  % For high order multiple system, we will discuss the systematics in mass-ratio in section~\ref{sec:multiplicity}.

Undiscriminating triple (or higher order multiple) system would not change the fraction of binaries, but may distort the mass-ratio distribution. The mass-ratio of the binary in this work is defined as the stellar mass of the secondary over the stellar mass of the primary, which is not the case for high-order multiplicity. In principle, without considering the tertiary contribution in luminosity, we may mistakenly attribute the total incremental luminosity of the secondary, tertiary and other high-order companions to a more massive secondary. Therefore, without discriminating high-order multiple system, the mass-ratio distribution may be systematically bias to larger $q$. 

We investigate 27 triple systems with spectral types or mass studied by \citet{Raghaven2010} and find that the mass of tertiary is much smaller than the secondary in 21 of them. Consequently, they would not significantly change the total luminosity. For instance, if the primary has mass of $0.7$\,\Msun, the mass of the secondary is $0.5$\,\Msun, then a tertiary with mass of $0.2$\,\Msun\ only makes \DMG\ brighter by $0.02$\,mag. When the mass of the tertiary increases to $0.4$\,\Msun, the \DMG\ increases by 0.08\,mag, roughly comparable to the uncertainty of \DMG. Therefore, when the mass of tertiary is significantly smaller than that of the secondary, its contribution in the mass-ratio distribution can be neglected. 

We also find 5 of the 27 triple systems, most of which contain brown dwarf companions, have the secondary and tertiary with similar masses, which can potentially affect the mass-ratio distribution. However, since these stars are only a small fraction in triple systems and triple systems occupy only a small fraction in multiplicities, their effect in the mass-ratio distribution is small and hence can be neglected. 

%%%%%%%%%%%%%%%%%%%%%%%%%%%%%%%%%%%%%%%%%%%%%%%%%%%%%%
\section{Conclusions}\label{sec:conclusion}
%%%%%%%%%%%%%%%%%%%%%%%%%%%%%%%%%%%%%%%%%%%%%%%%%%%%%%

In this work, we use more than 50\,000 solar-type MS stars selected from LAMOST and \gaia\ data to derive the binary fraction and mass-ratio distribution as functions of \feh\ and \mpri. We develop a hierarchical Bayesian model to simultaneously derive both the binary fraction and the mass-ratio distribution in each small metallicity and primary mass bin.

For the first time, we find that the binary fraction is tightly anti-correlated with $\gamma$, which is the power index of the power-law shape mass-ratio distribution. The data can be separated into two groups with different $\gamma$. We find that the \fbin\ of the \glow\ group with $\gamma<1.2$ displays clear correlation with \mpri\ but quite flat relationship with \feh. The $\gamma$ for this group, on the other hand, shows anti-correlations with both \feh\ and \mpri. For the \ghigh\ group with $\gamma>1.2$, \fbin\ shows significant anti-correlation with \feh\ but is not correlated with \mpri. Although $\gamma$ of the group is mildly correlated with \feh\ and \mpri, the large uncertainty prefers to that these correlations are not statistically substantial.

Moreover, looking at the distributions of \fbin\ and $\gamma$ in the \mpri\ vs. \feh\ plane, the two groups are quite different. The \ghigh\ group favorites smaller \mpri\ and larger \feh, resulting in a concentrated ridge-like regime in the \mpri\ vs. \feh\ plane, while the \glow\ group distributes in the regions with larger \mpri\ or lower metallicity. 

These complicated features strongly hint that the field binaries may experience efficient dynamical processing during the time when they still stayed in the embedded clusters in which they were born \citep{Kroupa1995a,Kroupa1995b}. As a result, the properties of the present-day binaries have been substantially reshaped, especially for those with lower primary mass and higher metallicity. 
Using the resulting \fbin\ and $\gamma$ at each \feh\ and \mpri\ bin as a prior, we are able to identify 5\,410 binary star candidates with $P_s/P_b<0.1$ from the 57\,978 samples. And we also derived the mass-ratio for all 57\,978 stars. 

We find that the number of samples is critical to derive reliable mass-ratio distribution. If the sample size is only a few hundreds, the derived mass-ratio distribution for the population with small binary fraction would systematically bias to high mass-ratio. However, when the sample size is large enough, such as larger than 2000, we can reliably work out both binary fraction and mass-ratio distribution based on the hierarchical Bayesian model.

\section*{Acknowledgements}
CL thanks Changqing Luo, Pavel Kroupa,  Zhanwen Han, Lu Li, Cheng-Yuan Li, Xiao-Bin Zhang, Licai Deng, and Haijun Tian for their helpful discussions and suggestions. This work is supported by the National Natural Science Foundation of China (NSFC) with grant No. 11835057. 
%CL is supported by the LAMOST Fellowship.
This project was developed in part at the 2018 NYC Gaia Sprint, hosted by the Center for Computational Astrophysics of the Flatiron Institute in New York City. 
It was also partly developed at 2018 Gaia-LAMOST sprint workshop supported by the NSFC under grants 11333003 and 11390372.
Guoshoujing Telescope (the Large Sky Area Multi-Object Fiber Spectroscopic Telescope LAMOST) is a National Major Scientific Project built by the Chinese Academy of Sciences. Funding for the project has been provided by the National Development and Reform Commission. LAMOST is operated and managed by the National Astronomical Observatories, Chinese Academy of Sciences.
This work has made use of data from the European Space Agency (ESA) mission {\it Gaia} (\url{https://www.cosmos.esa.int/gaia}), processed by the {\it Gaia} Data Processing and Analysis Consortium (DPAC, \url{https://www.cosmos.esa.int/web/gaia/dpac/consortium}). Funding for the DPAC has been provided by national institutions, in particular the institutions participating in the {\it Gaia} Multilateral Agreement.

%%%%%%%%%%%%%%%%%%%%%%%%%%%%%%%%%%%%%%%%%%%%%%%%%%



%\bibliographystyle{mnras}


%%%%%%%%%%%%%%%%%%%%%%%%%%%%%%%%%%%%%%%%%%%%%%%%%%

%%%%%%%%%%%%%%%%% APPENDICES %%%%%%%%%%%%%%%%%%%%%

\appendix

%%%%%%%%%%%%%%%%%%%%%%%%%%%%%%%%%%%%%%%%%%%%%%%%%%%%%%
\section{Initial stellar mass estimation}\label{sec:mass}
%%%%%%%%%%%%%%%%%%%%%%%%%%%%%%%%%%%%%%%%%%%%%%%%%%%%%%

\begin{figure}
	\centering
	\includegraphics[scale=0.5]{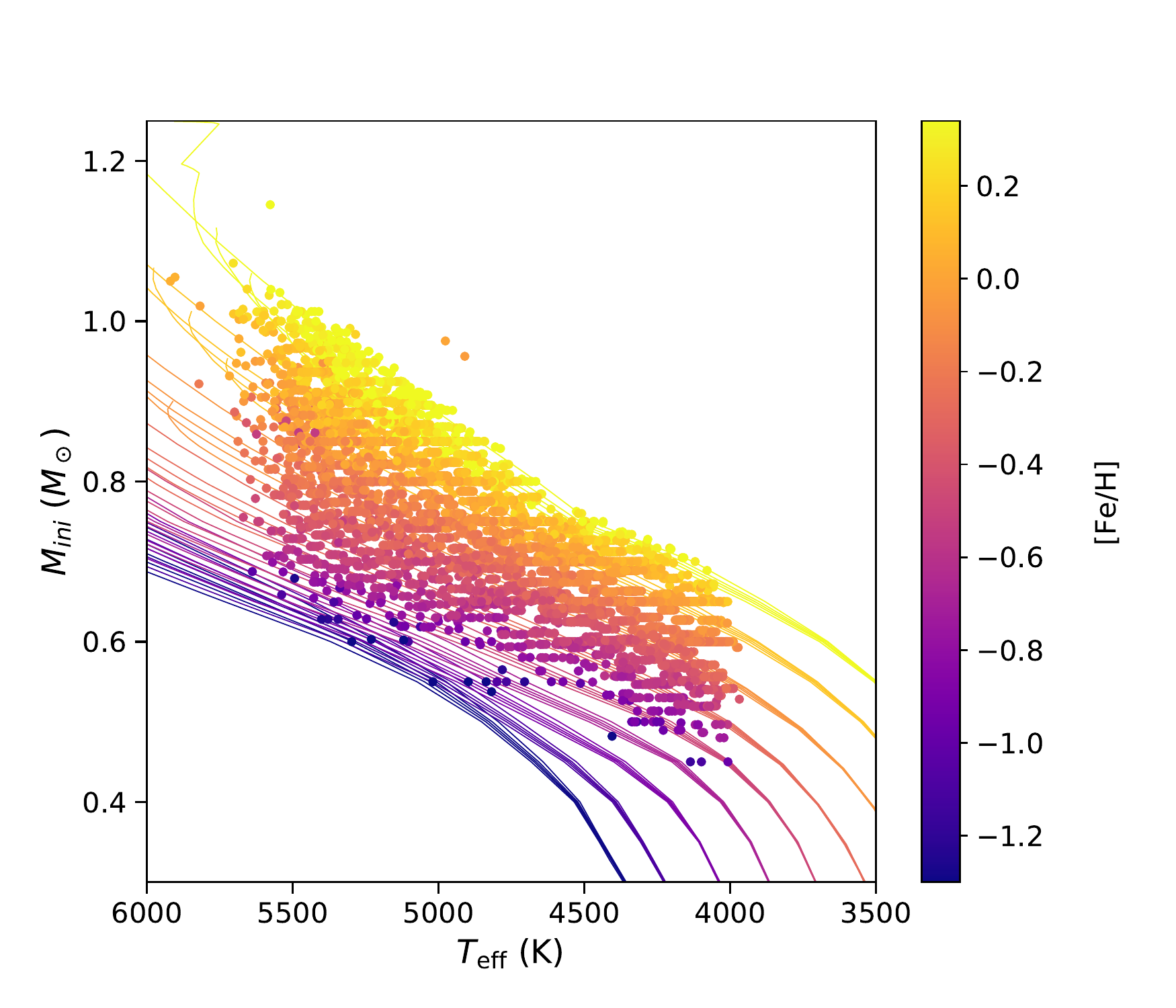}
	\caption{The locations of 10000 arbitrarily selected stars from the samples in \Mini\ vs. \teff\ plane with the color coded \feh. The color lines indicate the PARSEC isochrones with ages of 1, 3, 5, 7, 9, and 11\,Gyr at each metallicity value, which are coded with the same color as the stars.}\label{fig:Mass}
\end{figure}

The stellar mass of stars is determined by comparing the effective temperature (\teff), surface gravity (\logg), and metallicity (\feh) of stars with the PARSEC isochrones \citep{Bressan2012}. For the MS stars in this work, they are mostly not located in the star forming region and hence  should be older than 1 Gyr. Therefore, we remove the isochrones with age younger than 1\,Gyr so that the young isochrones would not affect the mass determination.

For each star, the initial stellar mass is determined by looking for the maximum likelihood in \teff, \logg, and \feh\ space. The likelihood distribution is written as
\begin{align}\label{eq:initialmass}
\ln\mathcal{L}\propto&-\frac{(T_{obs}-T_{model}(M_{ini},age)^2}{2\sigma_{T}^2}\nonumber\\
&-\frac{(G_{obs}-G_{model}(M_{ini},age))^2}{2\sigma_G^2}\nonumber\\
&-\frac{(Z_{obs}-Z_{model}(M_{ini},age))^2}{2\sigma_Z^2},
\end{align}
where $T$, $G$, and $Z$ are the effective temperature, surface gravity, and metallicity, respectively. The uncertainties of \teff, \logg, and \feh\ are $\sigma_T=110$\,K, $\sigma_G=0.2$\,dex, and $\sigma_Z=0.15$\,dex, according to \citet{Gao2015}. Based on \citet{ElBadry2018a}, we consider the averaged contribution of the unresolved secondaries in stellar parameter estimation and add systematic bias of 50\,K in \teff, uncertainty of 0.1\,dex in \logg, and uncertainty of 0.05\,dex in \feh\ to the total uncertainties and adopt $\sigma_T=120$\,K, $\sigma_G=0.22$\,dex, and $\sigma_Z=0.16$\,dex.

Because that the age for the late type MS stars is very difficult to be determined, we marginalize the age in Eq.~(\ref{eq:initialmass}) to derive the likelihood distribution of \Mini. We choose the value of \Mini\ at the maximum likelihood and determine the uncertainty from the 15\% and 85\% percentiles. The likelihood approach can reach typical precision of about 0.04--0.06\,\Msun\ for the stars with initial stellar mass between 0.4 and 1.0\,\Msun. For binary system, the stellar parameters are dominated by the primary stars. Therefore, we consider the mass estimates for the binaries as the mass of the primary.

Figure~\ref{fig:Mass} displays the locations of 10000 randomly selected MS stars from the samples in \Mini\ vs. \teff\ plane with the color coded \feh.% It is seen that the metallicity is more significantly affect the mass estimates than the age.

%\clearpage

%%%%%%%%%%%%%%%%%%%%%%%%%%%%%%%%%%%%%%%%%%%%%%%%%%%%%%
\section{Interstellar extinction correction}\label{sect:extinction}
%%%%%%%%%%%%%%%%%%%%%%%%%%%%%%%%%%%%%%%%%%%%%%%%%%%%%%
\begin{figure}
	\centering
	\includegraphics[scale=0.5]{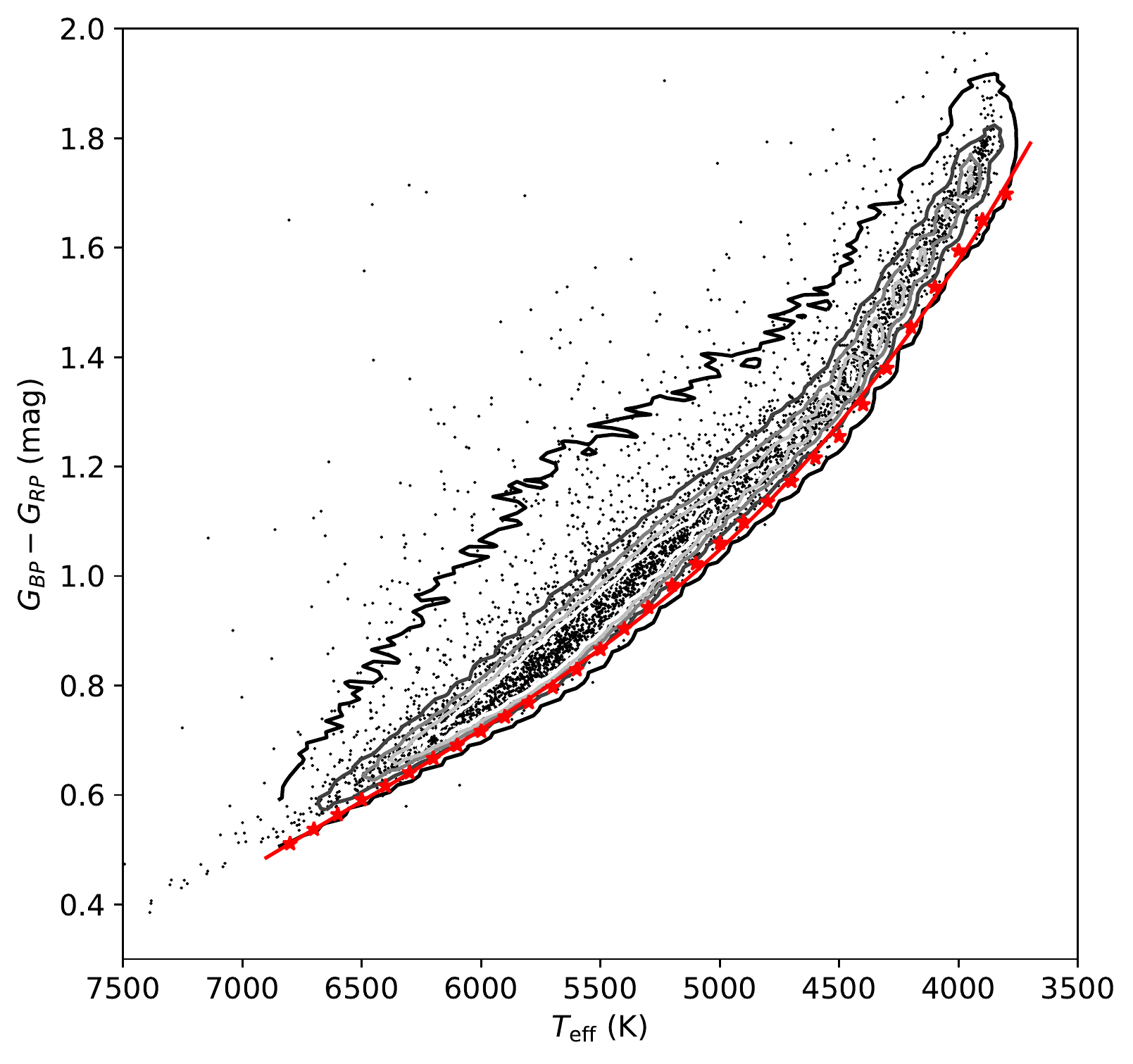}
	\caption{The small dots indicates the locations of the stars in our samples in the $G_{BP}-G_{RP}$ vs. \teff\ plane. The contours shows the distribution of these stars. The red stars indicate the 5\% percentiles, which are adopted as the intrinsic color indices, in the bluest end of $G_{BP}-G_{RP}$ at various \teff. The red solid line shows the best model of $(G_{BP}-G_{RP})_0$ as a cubic polynomial of \teff.}\label{fig:reddening}
\end{figure}
%To be more sensitive to the secondary star in a MS-MS binary system, we select the color index $G_{BP}-K_s$, in which $G_{BP}$ and $K_s$ are the $BP$ photometry in \gaia\ \citep{Gaia2018} and $K_s$ band photometry in 2MASS \citep{2mass}. 
We first derive the color excess of $G_{BP}-G_{RP}$ and then estimate the extinction of $G$ band from it with adopted extinction coefficients for $G$ band.
Figure~\ref{fig:reddening} shows the distribution of the sample stars in $G_{BP}-G_{RP}$ vs \teff\ plane. According to \citet{Ducati2001}, \citet{Wang2014}, and \citet{Xue2016}, the bluest edge at a given \teff\ is contributed by the stars with nearly no reddening in $G_{BP}-G_{RP}$. \citet{Jian2017} applied this approach and well determined the reddening of the LAMOST data. We follow the same method and adopt that the 5\% percentiles of $G_{BP}-G_{RP}$ at given \teff\ bins (red stars displayed in Figure~\ref{fig:reddening}) represent for the intrinsic color indices corresponding to \teff. Therefore, we can estimate the relationship by fitting the intrinsic colors with a cubic polynomial. We find the best-fit intrinsic color index is
%[-2.15880477e-11  4.22719163e-07 -3.01430054e-03  8.25019801e+00] [3.46495908e-12 5.51594875e-08 2.88452224e-04 4.95140085e-01]
\begin{align}\label{eq:reddening}
(G_{BP}-G_{RP})_0 =& 	-2.16(\pm0.35)\times10^{-11}T_{\rm eff}^3\nonumber\\
&+4.23(\pm0.55)\times10^{-7}T_{\rm eff}^2\nonumber\\
&-3.01(\pm0.29)\times10^{-3}T_{\rm eff}\nonumber\\
&+8.25(\pm0.50).
\end{align}

The extinction coefficients for \gaia\ bands are provided by \citet{Danielski2018} based on \gaia\ DR1 such as 
\begin{align}\label{eq:extinctcoeff}
k_X = &c_1+c_2(G_{BP}-G_{RP})_0+c_3(G_{BP}-G_{RP})_0^2+c_4(G_{BP}-G_{RP})_0^3\nonumber\\
&+c_5A_0+c_6A_0^2+c_7(G_{BP}-G_{RP})_0A_0,
\end{align}
where $k_X=A_X/A_0$. Given that $E(G_{BP}-G_{RP})=(k_{BP}-k_{RP})A_0$, we can solve Eq~(\ref{eq:extinctcoeff}) for $A_0$ by using coefficients, $c_1$, $c_2$, ..., $c_7$, corresponding to $G_{BP}$ and $G_{RP}$, respectively. And then $A_0$ is brought back to Eq~(\ref{eq:extinctcoeff}) with coefficients corresponding to $G$-band  to solve for $A_{G}$. We adopt the coefficients, $c_1$, $c_2$, ..., $c_7$ for $G_{BP}$, $G_{RP}$, and $G$, respectively, provided in Table 1 of \citet{HRD2018}.
 As a by-product, we can also obtain $A_{BP}$ and $A_{RP}$ by applying the same method. Adopting the extinction law of \citet*{Cardelli1989} with $R_V=3.1$, we can also estimate the color excess of $E(G-K_s)$ and subsequently give the intrinsic color index $(G-K_s)_0$.

It is noted that the estimation of $(G_{BP}-G_{RP})_0$ does not account for the effect induced by metallicity. We tried to derive the intrinsic color index for different metallicity and found that the difference of the intrinsic color index with metallicity is only a few hundredth magnitude. Therefore, we ignore the effect of metallicity in the intrinsic color index.

The uncertainty of the reddening correction is contributed from several channels. First, the cubic polynomial relation may affected by the uncertainties from the observed color index, which is usually as small as a few mill-magnitudes. Then, the uncertainty of \teff\ from the LAMOST pipeline, which is around 120\,K, is propagated to $E(G_{BP}-G_{RP})$ and contribute to about $0.04$\,mag. Finally, total uncertainty of $A_G$ is about $0.05$\,mag.

% Don't change these lines
\bsp	% typesetting comment
\label{lastpage}
\end{document}